\documentclass[aps,prb,twocolumn,superscriptaddress,showpacs,longbibliography,notitlepage]{revtex4-1}
\usepackage[colorlinks=true,citecolor=blue,linkcolor=blue,breaklinks=true]{hyperref}
\usepackage{amssymb}
\usepackage{graphicx}
\usepackage{amsmath}
\usepackage{mathrsfs}
\usepackage{mathtools}
\usepackage[export]{adjustbox}
\usepackage{epsfig}
\usepackage{times}
\usepackage{color}
\usepackage{setspace}
\usepackage{calc}
\usepackage{natbib}
\usepackage{bm}% bold math
\DeclareMathAlphabet{\mathpzc}{OT1}{pzc}{m}{it}
\usepackage{array}
\usepackage{multirow}
\begin{document}

%-----------------------------------------------------------------------------------------
\newcommand {\beq} {\begin{equation}}
\newcommand {\eeq} {\end{equation}}
\newcommand {\bqa} {\begin{eqnarray}}
\newcommand {\eqa} {\end{eqnarray}}
\newcommand{\dhat}{\ensuremath{\hat{D}}}
\newcommand{\ehat}{\ensuremath{\hat{E}}}
\newcommand{\lhat}{\ensuremath{\hat{\Lambda}}}
\newcommand{\zbar}{\ensuremath{\bar{\zeta}}}
\newcommand{\ebar}{\ensuremath{\bar{\eta}}}
\newcommand {\ba} {\ensuremath{b^\dagger}}
\newcommand {\Ma} {\ensuremath{M^\dagger}}
\newcommand {\psia} {\ensuremath{\psi^\dagger}}
\newcommand {\psita} {\ensuremath{\tilde{\psi}^\dagger}}
\newcommand{\lp} {\ensuremath{{\lambda '}}}
\newcommand{\A} {\ensuremath{{\bf A}}}
\newcommand{\Q} {\ensuremath{{\bf Q}}}
\newcommand{\kk} {\ensuremath{{\bf k}}}
\newcommand{\qq} {\ensuremath{{\bf q}}}
\newcommand{\kp} {\ensuremath{{\bf k'}}}
\newcommand{\rr} {\ensuremath{{\bf r}}}
\newcommand{\rp} {\ensuremath{{\bf r'}}}
\newcommand {\ep} {\ensuremath{\epsilon}}
\newcommand{\nbr} {\ensuremath{\langle ij \rangle}}
\newcommand {\no} {\nonumber}
\newcommand{\up} {\ensuremath{\uparrow}}
\newcommand{\dn} {\ensuremath{\downarrow}}
\newcommand{\rcol} {\textcolor{red}}
\newcommand{\bcol} {\textcolor{blue}}
\newcommand{\bu} {\bold{u}}
\newcommand{\tr}[1]{\mathrm{Tr}\left[#1\right]}
\newcommand{\ve}[1]{\boldsymbol{#1}}
\newcommand{\args}[1]{\ve{#1},\ve{\bar{#1}}}
\newcommand{\mes}[1]{\mathcal{D}\hspace{-2pt}\left[\args{#1}\right]}
\newcommand{\ii}{\iota}%\mathbf{i}
\newcommand{\mnm} {\ensuremath{\mathbb{M}}}
%-----------------------------------------------------------------------------------------

\begin{abstract}
We formulate a new ``Wigner characteristics'' based method to
calculate entanglement entropies of subsystems of Fermions using Keldysh
field theory. This bypasses the requirements of working with
complicated manifolds for calculating R\'{e}nyi entropies for many body
systems. We provide an exact
analytic formula for R\'{e}nyi and von-Neumann entanglement entropies of
non-interacting open quantum systems, which are initialised in arbitrary Fock
states. We use this formalism to look at entanglement entropies of
momentum Fock states of one-dimensional Fermions. We show that the entanglement
entropy of a Fock state can scale either logarithmically or linearly
with subsystem size, depending on whether the number of
discontinuities in the momentum distribution is smaller or larger than
the subsystem size. This classification of states in terms number of blocks
of occupied momenta allows us to analytically estimate the number of critical and
non-critical Fock states for a particular subsystem size. We also use
this formalism to describe entanglement dynamics of an open quantum
system starting with a single domain wall at the center of the
system. Using entanglement entropy and mutual information, we
understand the dynamics in terms of coherent motion of the domain wall
wavefronts, creation and annihilation of domain walls and incoherent exchange of
particles with the bath.

\end{abstract}
%-----------------------------------------------------------------------------------------
\title{Entanglement Entropy of Fermions from Wigner Functions: Excited
States and Open Quantum Systems}
\author{Saranyo Moitra}\email{smoitra@theory.tifr.res.in}
 \affiliation{Department of Theoretical Physics, Tata Institute of Fundamental
 Research, Mumbai 400005, India.}
\author{ Rajdeep Sensarma} 
\affiliation{Department of Theoretical Physics, Tata Institute of Fundamental
 Research, Mumbai 400005, India.}

\pacs{}
\date{\today}

\maketitle
%-----------------------------------------------------------------------------------------
\section{Introduction}

In a many body system, the erasure of information shows up as a
classical probability in the description of the system. For quantum
systems, if one starts from a generic
pure quantum state and traces out some degrees of freedom, the
description requires a ``reduced'' density matrix which allows quantum probability
amplitudes and classical probabilities to co-exist in a single
formalism\cite{HorodeckiReview}. Entanglement entropies (like Von-Neumann and R\'{e}nyi
entropies) are corresponding  entropy measures of this resulting
classical probability distribution, with the entanglement eigenvalues
(the eigenvalues of the density matrix) giving the classical
probability of finding the corresponding eigenstate in the
ensemble. In this sense, entanglement entropies measure how much
classical information is required to compensate for the loss of
knowledge due to tracing over degrees of freedom.

Entanglement has been extensively studied in the context of quantum
information and computation \cite{nielsen2010quantum}, and in recent
years, it has proved to be useful in many body physics as
well\cite{EEManybodyReview}. In particular, the scaling of bipartite
entanglement entropy with subsystem size in ground states\cite{EisertAreaLaws} has been used to detect topological phases\cite{jiang2012topological,KitaevPreskillTopo,Tarun_Ashvin_topo} and quantum phase transitions \cite{vidal2003entanglement}. More recently, finite energy density eigenstates have also been shown to exhibit some universal scaling of entanglement entropy(EE) \cite{RigolFreeFermions,RigolFreeFermionsXY,TGroverChaoticEigenstates}.
Recently, entanglement entropy of many body systems have been measured experimentally\cite{Islam2015_expt,lukin2019probing_expt}, although the system sizes are not in the thermodynamic limit.
However, theoretical understanding of the same is comparatively still limited and efficient computational methods capable of accessing large system sizes are restricted to non-interacting theories\cite{Peschel_2003} or special integrable models in one dimension.

There have been three main approaches to calculating bipartite entanglement
entropy of Fermionic quantum states: (1) Conformal Field Theory based
approaches, which have yielded strong and crisp analytic predictions for
size dependence of entanglement entropies in critical theories
~\cite{Calabrese_2009,BrianSwingle_CFT,FradkinMoore_CFT} (2)
Operator based approaches, which are confined to non-interacting
systems, but have yielded exact answers~\cite{Peschel_2003,Peschel_2009} and (3) Field theory based
approaches, which maps the problem to calculating partition
function of the system on a Riemann surface with replicated sheets,
with complicated boundary conditions \cite{CasiniHuerta,Goev2006,SubirO(N),SubirWitczakKrempa} . There have of course been numerical approaches, which try to
construct the state in the large dimensional Hilbert space either
exactly (exact diagonalization) ~\cite{yu2016bimodal_MBL,Abanin_EntanglementMBL,samanta2020extremal} or in approximate
ways ( DMRG ~\cite{HastingsMelko_QMC,TGrover_FermionsQMC}, Tensor
Networks~\cite{Hayes_TensorNetworks}%, PEPS~\cite{EEfromPEPS} etc
), and then work out the entanglement entropy by
constructing a Schmidt decomposition of the state. With improvement of
strategies to construct these states, this approach has yielded a
wealth of information about structure of entanglement entropy of
states. The numerical methods are less restricted than analytic ones
in terms of applicability, but suffer from the
problems of dealing with large Hilbert spaces.

In a recent paper, Chakraborty and Sensarma ~\cite{AhanaRajdeepBosons} showed that Wigner
characteristic functions for density
matrices of bosonic many body systems are equivalent to partition
functions in presence of time-localized sources (i.e. a delta function
kick).
The formalism can be adapted in a
straightforward way to reduced density matrices by restricting the
sources to the subregion where the reduced density matrix is supported. Combined with well known relations between Wigner functions and
R\'{e}nyi entropies of bosonic density matrices, this allowed a
calculation of R\'{e}nyi entropies of subsystems, which only involved correlation
functions of the original model without any complicated Riemann
surfaces. In this paper, we use similar ideas, combined with
construction of ``distribution functions'' for Fermions by
Glauber and Cahill~\cite{CahillGlauberF}, to show that one can construct Keldysh partition functions
of the fermionic systems with time localized sources living on a
subregion. One can use these Grassmann valued characteristic functions
and integrate over the sources to obtain the R\'{e}nyi entropies of
arbitrary order, and hence von-Neumann entropy for a system of Fermions, both in and out of
equilibrium. This part of the formalism was also developed independently by Haldar, Bera and Banerjee, who then went on to apply it to
SYK models~\cite{haldar2020renyi}.

In this paper, we focus our attention on the application of this
formalism to non equilibrium dynamics of Fermions in both open and
closed quantum systems, starting from arbitrary initial conditions. A
key result of this paper is that we obtain an exact analytic formula
for the entanglement entropy of an open quantum system of Fermions
whose dynamics start from an arbitrary initial Fock state. To our
knowledge, this is the first time such a formula has been derived for
open quantum dynamics with arbitrary initial conditions.

While a large amount of work has gone into understanding the
entanglement properties of ground states of
Fermions~\cite{EisertAreaLaws}, the excited states have received
relatively less analytic attention. Early work in this direction by Alba et. al ~\cite{Alba_2009}
on entanglement entropy of excited states of $XY$ model showed the
presence of ``critical'' states, whose entanglement entropy varied
logarithmically with the size of the subsystem. Later
works by Ares et.al{~\cite{Ares_2014}, Storms et al.~\cite{storms2014entanglement}, and more
 recently the works of Vidmar et al.~\cite{RigolFreeFermions,RigolFreeFermionsXY}, Carrasco
  et al.~\cite{Carrasco2017}, Jafarizadeh et al.~\cite{Rajabpour_excitedstates}, and Lu et al.~\cite{Tarun_chaoticeigenstates} have
  revived interest in the entanglement entropy of Fock states of Fermions
We apply our exact formula to study the
entanglement entropy of arbitrary momentum Fock states of spinless
Fermions in one dimension. Here, our main results are: (i) For studying the behaviour of
entanglement entropy, it is useful to classify Fock states in
terms of number of blocks of occupied momenta in that state. This is
similar to the classification proposed by Alba and
Calabrese~\cite{Alba_2009} (ii) The entanglement entropy of a given
Fock state scales logarithmically (``critical'' behaviour) with the
subsystem size if the subsystem size is much larger than the number of
blocks of occupied momenta (or number of discontinuities in the momentum distribution) of the Fock state. The
entropy scales linearly (``non-critical'' behaviour) with the
subsystem size when the number of discontinuities is much larger than the subsystem
size. Thus Fock states are neither critical or non-critical by
themselves, the same Fock state can appear critical or non-critical
depending on the range of subsystem sizes one is probing. (iii) For a
given subsystem size, we obtain an
analytic estimate of the number of critical states in terms of the
density of Fermions.

Entanglement dynamics in open quantum systems have been studied within
the master equation approach~\cite{Benatti_2010}  and quantum trajectory approach~\cite{Nha_Carmichael_2004} , but they have been restricted to systems with $2$ or
few qubits\cite{Aolita_2015_OQSEntanglement}. A general analytic treatment of entanglement dynamics of many body
systems starting from different initial conditions has been missing in
the literature. Our exact formula fills this void for non-interacting
Fermionic systems. Using this, we study the dynamics of an open
quantum system of one dimensional spinless Fermions connected to a
bath. We initialize this system in a Fock state
with a density profile where the left half of the lattice is occupied and the right half is empty, creating
a domain wall in the chain. We measure the time
evolution of entanglement entropy of subsystems of different sizes
placed at different locations in the system. We also look at the
mutual information between the different components of a subsystem to
see how they are entangled between themselves. We find that the
dynamics of entanglement can be understood in terms three different
processes: (i) an incoherent exchange of particles with the bath,
which leads to a background evolution of entanglement, which does not
depend on the location of the subsystem. (ii) A coherent wave of
domain wall propagating and reflecting off the boundaries of the
system, which leads to sharp jumps in entanglement entropy. These jumps occur at different times for subsystems at
different locations, revealing the propagation and reflection of the
wave. The coherent wave is damped by dissipation from the bath, and
(iii) Local Rabi oscillations between nearest neighbours which also
lead to splitting and merging of domains. This process leaves its
imprint in the form of oscillations superposed on an overall
background. We show that once the coherent wave hits the subsystem
there is a sharp increase in the mutual information between the
components of the subsystem, showing that they are getting entangled
in the process. Finally, as the system evolves from a pure quantum
state to a density matrix characterized by a thermal ensemble in the
long time limit, we focus on the additional entropy density (entropy
per site) of a
subsystem vis-a-vis the full system, which represents the entropy of
information loss due to tracing of degrees of freedom. We show that
while the time evolution of entanglement entropy of larger subsystems closely follow that of the full system, the effects of the quantum
processes can be cleanly demonstrated in the evolution of the excess
entropy density.

We now provide a roadmap for the different sections of the paper: (i)
In Section \ref{wfn_Rny}, we define the Grassmann valued ``Wigner
characteristic function'' of a fermionic density matrix, previously
discussed by Glauber and Cahill~\cite{CahillGlauberF}, and show how one can relate integrals
(over Grassmann valued arguments) of these functions to
R\'{e}nyi entropies of different orders. (ii)  In Section \ref{WignerKeldysh},
we show how the Keldysh partition function of a Fermionic system with
particular arrangement of sources is equal to the Wigner
characteristic function of the reduced density matrix of a
subsystem.
%This part is applicable to general interacting models both in and out
%of equilibrium.
(iii) In Section \ref{ArbInit}, we will extend this
formalism to the case of non-equilibrium dynamics starting from
arbitrary initial states. (iv) In section \ref{ExactFormula}, we
derive an exact analytic formula for R\'{e}nyi and von Neumann entropies
of a subsystem of a open quantum system of non-interacting Fermions
starting from an arbitrary Fock state.
%An alternative derivation
%leading to a diagrammatic evaluation of entanglement entropy is given
%in Appendix \ref{diag}
(vi) In section \ref{Fock}, we
will focus our attention on arbitrary momentum Fock states and show
that they can exibit linear
or logarithmic scaling of entanglement entropy with subsystem size,
depending the range of subsystem sizes one is probing. (vii) In section \ref{OQS}, we
will study the evolution of entanglement entropy and mutual
information in an open quantum system of one dimensional Fermions starting from a state with one half occupied and the other half empty. 

\section{ Wigner Functions and Entanglement Entropy of
  Fermions \label{wfn_Rny}}

We will study a system of $N$ spinless Fermions on a lattice with $V$
sites and calculate R\'{e}nyi and von Neumann entropies of a
subsystem $A$ with $V_A$ sites. We
will consider a complete single particle basis $i$ with $V$ distinct
values, and corresponding creation/annihilation operators
$c^\dagger_i$ ( $c_i$).
%Here $i$ can be
% a spatial index or a momentum label or
% any complete single-particle basis with $V$ distinct values.
The Fermionic coherent states,
defined by $|\ve{\zeta}, \ve{\bar{\zeta}}\rangle:=\dhat(\ve{\zeta},
\ve{\bar{\zeta}})|0\rangle$, are labelled by tuples of Grassman
variables $\ve{\zeta}=(\zeta_1,\zeta_2,\mathellipsis,\zeta_M)^{T}$ and
$\ve{\bar{\zeta}}=(\zbar_1,\zbar_2,\mathellipsis,\zbar_M)$, where $M$ is the number of modes, 
$|0\rangle$ is the vacuum state, and the displacement operator is given
by
\beq
\dhat(\args{\zeta}):=
\mathrm{e}^{\sum_i
  c^\dagger_i\zeta_i-\bar{\zeta}_ic_i}=\prod_i[1+\zeta_ic^\dagger_i-\bar{\zeta}_ic^\dagger_i-\bar{\zeta}_i\zeta_i(1/2-c^\dagger_ic_i)]
\eeq
These operators can be combined using
\beq
\dhat(\args{\zeta})\dhat(\args{\eta})=\dhat(\ve{\zeta}+\ve{\eta},\ve{\bar{\zeta}}+\ve{\bar{\eta}})\,\mathrm{e}^{\frac{1}{2}\left(\ve{\bar{\eta}}\cdot
  \ve{\zeta}-\ve{\bar{\zeta}}\cdot\ve{\eta}\right)}.
\eeq
To construct various ``quasi-distribution functions'', in a vein
similar to the Wigner distribution for Bosons\cite{CahillGlauberB}, it is useful to
introduce another operator $\ehat(\args{\zeta})$ where
\beq
\ehat(\args{\zeta})=\prod_i(1-2c^\dagger_ic_i)\dhat(\args{\zeta})=\mathrm{e}^{i\pi \hat{N}_{tot}}\dhat(\args{\zeta}).
\eeq
$\hat{N}_{tot}$ is the total Fermion number operator. $\ehat$
is thus related to $\dhat$ by the fermion parity
operator. This operator was first introduced
by Cahill and Glauber ~\cite{CahillGlauberF}, although our definition
differs from theirs by a sign of the arguments. 

In this case, one can easily show that the delta function over Grassmans\footnote{A delta function over grassmans satisfies $f(\alpha)=\int d\bar{\gamma}d\gamma \delta(\alpha-\gamma)f(\gamma)\quad \forall f$} can be represented as
\beq
\delta(\ve{\zeta}-\ve{\eta}):=\prod_i
(\zeta_i-\eta_i)( \bar{\zeta}_i- \bar{\eta}_i)=\tr{\dhat(\args{\zeta}) \ehat(\args{\eta})}
\label{trace_id}
\eeq
where $\eta_i$ and $\bar{\eta}_i$ are Grassman variables. We note that
any operator in the Fermionic Fock space, which preserves total
Fermion parity, can be expanded in terms of
either $\dhat$ or $\ehat$ as
\bqa
\no \hat{F} &=& \int \mes{\zeta} f_D(\args{\zeta})\dhat(\args{\zeta})\\
&=&  \int \mes{\zeta} f_E(\args{\zeta})\ehat(\args{\zeta})
\label{opexp}
\eqa
where $\mes{\zeta}:=\prod_i \mathrm{d}\bar{\zeta}_i\mathrm{d}\zeta_i$
and 
\begin{align}
	\no f_D(\args{\zeta}) &= \tr{\hat{F}\ehat(\args{\zeta})}\\
	f_E(\args{\zeta})&= \tr{\dhat(\args{\zeta})\hat{F}} 
	\label{opexpcoeff}
\end{align}
As a consequence of this, one can expand $\ehat$ in terms of $\dhat$,
\begin{equation}
\ehat(\args{\zeta})=2^{M}\int \mes{\eta} \dhat(\args{\eta})\mathrm{e}^{\frac{1}{2}\sum_i\bar{\zeta}_i\eta_i-\bar{\eta}_i\zeta_i}
\end{equation}
One can now define the equivalent of the Wigner characteristic function
for fermions,
\begin{align}
	\chi_D(\args{\zeta})&= \rho_E(\args{\zeta})=%
	\tr{\hat{\rho}\dhat(\args{\zeta})} \\
	\no \chi_E(\args{\zeta})&= \rho_D(\args{\zeta})=%
	\tr{\hat{\rho}\ehat(\args{\zeta})} \\
	\no &=2^{M}\int \mes{\eta} \chi_D(\args{\eta})\mathrm{e}^{\frac{1}{2}\sum_i\bar{\zeta}_i\eta_i-\bar{\eta}_i\zeta_i}
\end{align}
 where $\hat{\rho}$ is the density matrix of the system in a $2^M$ dimensional Hilbert space. For a reduced
 density matrix of a subsystem of $V_A$ sites, $M=V_A$.
 %While one can define a ``Wigner function
%equivalent'' for Fermions, this is
%a Grassmann valued function of Grassmann variables and, unlike Bosonic Wigner functions, does not
%correspond to anything measurable.
 All operator expectations can be
written in terms of $\chi_D$ and we will later see that $\chi_D$ can be
calculated within a path-integral/field theoretic approach.
%Thus we
%will keep working with the characteristic functions defined above.

Using the expansions, Eq.~\ref{opexp} and
Eq.~\ref{opexpcoeff}, together with the identity
Eq.~\ref{trace_id}, the expectation of parity preserving operators are
given by
\beq
\langle \hat{F}\rangle =\tr{\hat{F}\hat{\rho}}=\int {\cal
  D}[\args{\zeta}]~f_D(\args{\zeta})~\chi_D(\args{\zeta})
\eeq
The second R\'{e}nyi entropy, $S^{(2)}= -\ln\tr{\hat{\rho}_r^2}$, is
\begin{widetext}
\beq
S^{(2)}=-\ln\left[ 2^{V_A}\int\mes{\zeta}\mes{\eta}\hspace{1pt}%
\chi^r_D(\args{\zeta})\, \chi^r_D(\args{\eta})\,
\mathrm{e}^{\frac{1}{2}\sum^{'}_x\bar{\zeta}_x\eta_x-\bar{\eta}_x\zeta_x}\right]
\eeq
\end{widetext}
where $\hat{\rho}_r$ is the reduced density matrix. Here the $'$ in
the summation indicates that the spatial index $x$ runs only over the subsystem
$A$. Note that in case of continuum theories, the sum will be
replaced by appropriate integrals. The above result can be generalized
to the trace of $n$ operators to get the relation between the $n^{th}$
order R\'{e}nyi entropy $S^{(n)}$ and the Wigner characteristic of the
reduced density matrix
\begin{widetext}
  \begin{equation}
  	\begin{split}
  S^{(n)}=\frac{1}{1-n} \ln \left[ 2^{({n-1})V_A}%
  \int  \prod_{i=1}^{n-1}%
  	\mathcal{D}[\ve{\zeta}^{(i)},\ve{\zbar}^{(i)}]%
  	\mathcal{D}[\ve{\eta}^{(i)},\ve{\bar{\eta}}^{(i)}]%
  	%\left[\begin{array}{c}
  	%\ve{\zeta}^{(i)},\ve{\zbar}^{(i)}\\%
  	%\ve{\eta}^{(i)},\ve{\bar{\eta}}^{(i)}%
	%\end{array}\right]
   \prod_{i=1}^{n-1}
   \chi^r_D[\ve{\eta}^{(i)},\ve{\bar{\eta}}^{(i)}]%
   ~\chi^r_D\left[{\small \sum}\ve{\zeta}^{(i)},{\small \sum}\ve{\zbar}^{(i)}\right]\right.\\%
   \left.\exp\frac{1}{2}\left({\sum_{i}\ve{\zbar}^{(i)}\!\cdot\ve{\eta}^{(i)}-\ve{\bar{\eta}}^{(i)}\!\cdot\ve{\zeta}^{(i)}%
   +\sum_{i>j}\ve{\zbar}^{(i)}\!\cdot\ve{\zeta}^{(j)}-\ve{\bar{\zeta}}^{(j)}\!\cdot\ve{\zeta}^{(i)}}\right)
   \right]
   	\end{split}\label{eqref:Sngen}
  \end{equation}
\end{widetext}
We note that if all the Renyi entropies are analytically known, the von Neumann entropy can be calculated by analytic continuation $ S_{vN} := \lim_{n \rightarrow 1} S^{(n)}$. Having related all the R\'{e}nyi entanglement entropies of a subsystem of
fermions to the Wigner characteristic function of the reduced density
matrix $\chi^r_D$, we now focus our attention on methods to compute
this function. In the next section, we will relate $\chi_D^r$ to a
Schwinger Keldysh partition function of the fermionic system in
presence of a particular set of sources.

\section{Keldysh Field Theory and Wigner Characteristic Function\label{WignerKeldysh}}
Schwinger Keldysh field theories can describe quantum dynamics of both
open and closed quantum many body systems out of thermal
equilibrium. The key idea is to consider the time evolution of the
density matrix,
$\hat{\rho}(t)=\hat{U}(t,0)\hat{\rho}(0)\hat{U}^\dagger(t,0)$ and
  expand the forward time evolution operator $U$ in a path/functional
  integral with Grassmann fields $\psi_+(x,t)$ and
  $\bar{\psi}_+(x,t)$. The backward time evolution operator has
  a similar expansion in terms of $\psi_-(x,t)$ and
  $\bar{\psi}_-(x,t)$. This leads to a field theory with two copies of
  fields at each space time points.

  Observables $O(t) = \tr{\hat{O}\hat{\rho}(t)}/tr{\hat{\rho}(t)}$ are usually obtained in the field theoretic formalism by
  coupling sources to the fields, $J_{\pm}(x,t)$ coupling to
  $\bar{\psi}_\pm(x,t)$ and $\bar{J}_{\pm}(x,t)$ coupling to
  $\psi_\pm(x,t)$, and taking derivatives with respect to these
  sources, before setting the sources to zero. For time local
  observables, the standard practice is to take a symmetric linear
  combination of placing the operator on the $+$ and $-$ contour;
  i.e. $O(t)=[O_+(t)+O_-(t)]/2$.

  It was first shown in Ref~\onlinecite{AhanaRajdeepBosons}, that the
  expectation of the displacement operator for a system of Bosons is
  equivalent to the Keldysh partition function in presence of a
  particular set of sources. We follow similar algebra to show that
  this holds for Fermionic Wigner characteristics as well. 
  %(this aspect was \rcol{jointly developed} with Ref~\onlinecite{haldar2020R\'{e}nyi}). 
  The key innovation in this derivation is to consider
  \begin{widetext}
  \beq
\chi_D(\args{\zeta}, t)=\frac{\tr{\hat{U}(\infty,t)
    \hat{D}^{1/2}(\ve{\zeta}/2,\bar{\ve{\zeta}}/2)\hat{U} (t,0)\hat{\rho}_0 \hat{U}^\dagger(t,0)
    \hat{D}^{1/2}(\ve{\zeta}/2,\bar{\ve{\zeta}}/2)\hat{U}^\dagger(\infty,t)}}{\tr{ \hat{U} (\infty,0) \hat{\rho}_0 \hat{U}^\dagger(\infty,0)}}
\eeq
\end{widetext}
 or $D \sim D^{1/2}_+D^{1/2}_-$,  i.e. a multiplicative rather than a
 linear decomposition. We note
  that $\hat{D}^{1/2}$ is not a normal ordered operator. However,
  considering the anti-commutation of the Grassman fields, it can be
  shown that the insertion of  $\hat{D}^{1/2}$ on the $\pm$ contour is
    equivalent to turning on a source $J_\pm(x,\tau) = \pm \ii
    \zeta_x \delta(\tau-t)$. We note that the change of sign
    between the sources on the $+$ and $-$ contour is due to the fact
    that the action on the $-$ contour has a $-$ sign relative to the
    action on the $+$ contour in the Keldysh formalism. Thus we can
    identify
    \begin{widetext}
     \beq
\chi_D(\args{\zeta}, t)=Z\left[J_\pm(x,\tau) = \pm \ii
    \zeta_x \delta(\tau-t),\bar{ J}_\pm(x,\tau) = \mp \ii
   \bar{ \zeta}_x \delta(\tau-t)\right]
    \eeq
    
i.e. Wigner characteristic is the partition function with the above
set of sources. For fermionic systems, it is useful to work with symmetric and antisymmetric combinations of the fields,
  \begin{equation}
  	\begin{aligned}
  \psi_1(x,t)&=[\psi_+(x,t)+\psi_-(x,t)]/\sqrt{2}&\quad%
  \psi_2(x,t)&=[\psi_+(x,t)-\psi_-(x,t)]/\sqrt{2}\\
  \bar{\psi}_1(x,t)&=[\bar{\psi}_+(x,t)-\bar{\psi}_-(x,t)]/\sqrt{2}&\quad
  \bar{\psi}_2(x,t)&=[\bar{\psi}_+(x,t)+\bar{\psi}_-(x,t)]/\sqrt{2}.
  	\end{aligned}
  \end{equation} 
  One can similarly define sources  
  $J_1(x,t)=[J_+(x,t)+J_-(x,t)]/\sqrt{2}$,
  $J_2(x,t)=[J_+(x,t)-J_-(x,t)]/\sqrt{2}$,
  $\bar{J}_1(x,t)=[\bar{J}_+(x,t)-\bar{J}_-(x,t)]/\sqrt{2}$
  and
  $\bar{J}_2(x,t)=[\bar{J}_+(x,t)+\bar{J}_-(x,t)]/\sqrt{2}$, which
  couple to the respective rotated fields. If one considers the source
  pattern required for evaluating the Wigner characteristic in this Keldysh rotated
  basis, one finds that $J_1(x,\tau)=\bar{J}_2(x,\tau)=0$, with 
  $\bar{J}_1(x,\tau)=
  -\frac{\ii}{\sqrt{2}}\bar{\zeta}_x\delta(t-\tau)$ and $J_2(x,\tau)=
  \frac{\ii}{\sqrt{2}}\zeta_x\delta(t-\tau)$. This leads to the
  final result
    \beq
\chi_D(\args{\zeta}, t)=Z\left[J_1(x,\tau)=0,\bar{J}_1(x,\tau)=
  -\frac{\ii}{\sqrt{2}}\bar{\zeta}_x\delta(t-\tau), J_2(x,\tau)=
  \frac{\ii}{\sqrt{2}}\zeta_x\delta(t-\tau), \bar{J}_2(x,\tau)=0 \right]
  \eeq
  \end{widetext}
Note that the above equation is true for general interacting open or
closed systems of Fermions.

To calculate R\'{e}nyi entropies, one divides the system into subsystems
$A$ and $B$ and traces the density matrix over degrees of freedom
residing in $B$ to obtain a reduced density matrix. One then considers
traces of products of such reduced density matrices. The standard
field theoretic way of calculating this is to consider a field theory
on a replicated manifold with complicated boundary conditions on the
fields in the region $A$. In our formalism, the calculation of
partition function naturally traces over degrees of freedom. To
calculate the Wigner characteristic of the reduced density matrix
(instead of the full density matrix), we
simply need to restrict the sources to be nonzero in the region $A$
rather than the whole system. This makes our formalism ideally suited
for calculating entanglement entropies of subsystem of Fermions.

% Let us consider a simple example of $N$ non-interacting Fermions on a
% lattice of $L$ sites. For, now, we will consider spinless Fermions so
% that the total number of modes (in any basis) is $L$. Our aim is to
% calculate the R\'{e}nyi entropy of a subsystem of $L_A$ sites. We will
% work with finite $N$, $L$ and $L_A$ and carefully take thermodynamic
% limit, where $N,L,L_A \rightarrow \infty$, such that the density
% $N/L={\cal \rho}$ is fixed to a constant value. The case of subsystem
% fraction $f=L_A/L \rightarrow 0$ and $f\rightarrow constant$ will be
% treated separately.

We note that the standard Keldysh Field theory can treat non-equilibrium dynamics of the system (including coupling to baths), provided the initial state is in thermal equilibrium. We will deal with the case of arbitrary initial conditions in the next section. For thermal initial states, the gaussian action for a non-interacting system can be written as
\beq
S= \int dt \int dt'\sum_{x,x'} \psi^\dagger(x,t) G^{-1}(x,t;x',t') \psi(x',t')
\eeq
Here $\psi^\dagger(x,t)= [\bar{\psi}_1(x,t),\bar{\psi}_2(x,t)]$. The inverse propagator $G^{-1}$ and the one particle Green's functions
$G$ have the structure
%\begin{widetext}
  \begin{equation}
  \hat{G}^{-1} =\left[
  \begin{array}{cc}
	(\hat{G}^R)^{-1} &  (\hat{G}^{-1})^{K}\\	
	0 & (\hat{G}^A)^{-1}
    \end{array}\right] ~~~~~ \hat{G} =\left[\begin{array}{cc}%
                            \hat{G}^R &  \hat{G}^K\\
                  0 & \hat{G}^A
                      \end{array}\right]
\end{equation}
% \end{widetext}
where $\hat{G}^{R(A)}$ is the retarded (advanced) one particle Green’s function, and $\hat{G}^K$ is the Keldysh Green’s function .  The equal time Keldysh Green’s function is related to the physical one particle correlators in the system. In this case, the functional integrals over the fermion fields can be
carried out to compute the Wigner characteristic of the reduced
density matrix
\beq
\chi^r_D(\args{\zeta} ,t)= \mathrm{e}^{-\frac{1}{2} \sum^{'}_{xx'}\bar{\zeta}_x [\ii G^K(x,t;x',t)]\zeta_{x'}}
\eeq
where once again $x$ and $x'$ are restricted to the subregion $A$, and
we have used the fact that partition function of a Keldysh field
theory in absence of external sources is $1$ (this takes care of a
factor of $\det\hat{G}^{-1}$ coming from the functional integrals). We will
not comment on the particular form of $\hat{G}^K$ here, except reminding the
readers that $\hat{G}^K$ is an anti-hermitian matrix in the space-time
indices. This form of $\chi_D$ can then be used to calculate the
second R\'{e}nyi entropy
\begin{widetext}
\begin{equation}
S^{(2)}(t) = -\ln\left[2^{L_A}\hspace{-2pt}\int {\cal D}[\args{\zeta}]{\cal
    D}[\args{\eta}]~\mathrm{e}^{-\frac{1}{2}
    \sum^{'}_{xx'}(\bar{\zeta}_x,\bar{\eta}_x)\left(\begin{array}{cc}
                                       \ii G^K(x,t;x',t) & - \delta_{x,x'}\\
                                      \delta_{x,x'} & \ii G^K(x,t;x',t) \end{array}\right) \left(\begin{array}{c}
                                       \zeta_{x'}\\
                                       \eta_{x'} \end{array}\right)}\right]
                               =-\tr{\ln [\hat{1}+(\ii\hat{G}^K(t))^2]}
                             \end{equation}
where the matrix $\hat{G}^K(t)=G^K(x,t;x',t)$ is in the space of spatial
co-ordinates running over subregion $A$. One can in fact write down a general expression for the
$n^{th}$ order R\'{e}nyi entropy of the system in terms of the Keldysh
Green's function of the system
\begin{equation}
S^{(n)}(t) =\frac{1}{1-n} \tr{\ln\left[\left(\frac{\hat{1}-\ii \hat{G}^{K}(t)}{2}\right)^n\hspace{-5pt}+\left(\frac{\hat{1}+\ii \hat{G}^{K}(t)}{2}\right)^n\right]}
\end{equation}
 We can analytically continue the expression for $S^{(n)}$ to $n\to1$ to get 
 \begin{equation}
	S_{\text{vN}}=-\mathrm{tr}\left[\ln\left[ \left(\frac{\hat{1}-\ii \hat{G}^{K}(t)}{2}\right)	\ln\left(\frac{\hat{1}-\ii \hat{G}^{K}(t)}{2}\right)+ \left(\frac{\hat{1}+\ii \hat{G}^{K}(t)}{2}\right)\ln\left(\frac{\hat{1}+\ii \hat{G}^{K}(t)}{2}\right)\right]\right]
 \end{equation}
This recovers the well known formula of Ref\onlinecite{Peschel_2003} with the role of the correlation matrix played by $({\hat{1}-\ii \hat{G}^{K}(t)})/{2}$
\end{widetext}

\section{Arbitrary Initial conditions\label{ArbInit}}
A large class of interesting problems regarding dynamics of
entanglement entropies require description of dynamics starting from
non-thermal initial states; e.g. we may be interested in starting an
open quantum 
system in a product state in real space (with zero R\'{e}nyi entropy) and
describe the growth of entanglement as the system thermalizes. While
textbook Keldysh field theory requires a thermal initial state, recent developments~\cite{AhanaRajdeepArbitInit}  have provided a way to describe quantum
dynamics of many body systems starting from arbitrary initial
conditions.% (earlier textbook versions required thermal initial conditions)
By focussing at the initial time, the formalism can
be easily adapted to investigate the
entanglement entropy of particular Fock states or density matrices. 

Here we will briefly review the formalism of incorporating initial
athermal states as applied to the
calculation of the Wigner characteristics\cite{AhanaRajdeepArbitInit}. Consider an initial density
matrix of the form $\hat{\rho}_0=\sum_{\{n\}}c_{\{n\}}| \{n\}\rangle\langle
\{n\}|$ where $|
\{n\}\rangle=\otimes_\alpha |n_\alpha\rangle$ is the occupation number state, and $\alpha$ denotes a
single particle basis state. Note that $\alpha$ does not need to be a
spatial co-ordinate; this formalism will work for any complete one particle basis. In this case, we need to add to the
original Keldysh action a bilinear source term at $t=0$, $\delta S(u) =\ii\sum_\alpha 
\psi^\ast_1(\alpha,0)\psi_2(\alpha,0)\frac{1-u_\alpha}{1+u_\alpha}$. One
then calculates the Wigner characteristic in this theory by adding
appropriate sources and calculating the partition function in presence
of these sources,
\begin{widetext}
    \begin{equation}
    \chi^r_D(\args{\zeta}, t|\ve{u})=Z\left[%
    J_1(x,\tau)=0,%
    \bar{J}_1(x,\tau)=-\frac{\ii}{\sqrt{2}}\bar{\zeta}_x\delta(t-\tau), J_2(x,\tau)=
  \frac{\ii}{\sqrt{2}}\zeta_x\delta(t-\tau),
  \bar{J}_2(x,\tau)=0, u \right]
  \end{equation}
  \end{widetext}
The Wigner characteristic for the dynamics with the initial condition
is then given by
\beq
\chi^r_D(\args{\zeta}, t,) ={\cal L}(\partial_u,\rho_0) {\cal N}(\ve{u}) \chi_D(\args{\zeta}, t|\ve{u})\vert_{u=0}
\eeq
where ${\cal N}(u) =\prod_\alpha (1+u_\alpha)$ and ${\cal  L}=\sum_{\{n\}}c_{\{n\}}
\prod_{\alpha \in \mathcal{A}} \partial_{u_\alpha}$. ${\mathcal
  A}$ is the set of occupied modes in $|\{n\}\rangle$. For the case of
an initial pure Fock state,
we get 
\begin{equation}
\chi^r_D(\args{\zeta}, t) =\prod_\alpha [1+n_\alpha
\partial_{u_\alpha}] \chi_D(\args{\zeta}, t|\ve{u})\vert_{u=0}
\label{Wigchar_rho}
\end{equation}
Let us understand the consequences of these equations in the case of a
non-interacting theory (with coupling to baths). The Wigner
characteristic in presence of the sources $u$ will be given by
\begin{equation}
\chi^r_D(\args{\zeta}, t|\ve{u})=\exp\left[{-\frac{1}{2} \sum_{x,x'\in A}\zbar_x [\ii G^K(x,t;x',t|\ve{u})] \zeta_{x'}}\right]\label{chidu}
\end{equation}
In this
case, the Keldysh Green's function in presence of the $u$ sources is given
by
\begin{widetext}
	\begin{equation}
		\ii G^K(x,t;x',t'|\ve{u}) =%
		 \sum_\alpha
		G^R(x,t;\alpha,0)G^A(\alpha,0;x',t')\frac{1-u_\alpha}{1+u_\alpha}%
		+\ii\int_0^t\hspace{-5pt}dt_1\hspace{-2pt}\int_0^{t'}\hspace{-5pt}
		dt_2 G^R(x,t;x_1,t_1)\Sigma^K(x_1,t_1;x_2,t_2)G^A(x_2,t_2;x',t')
		\label{gku}
	\end{equation}

where $\Sigma^K$ is the Keldysh self energy due to possible coupling
to external baths. Note once again that $\alpha$ is not necessarily a
spatial co-ordinate; it can for example denote momentum labels.
The physical Keldysh Greens function $\mathcal{G}^{K}$ is given in terms of the $\ve{u}$ dependent ones as
	\begin{align}
		\ii\mathcal{G}^{K}(x,t;x',t')&={\cal L}(\partial_u,\hat{\rho}_0)[ {\cal N}(\ve{u})\ii G^{K}(x,t;x',t'|\ve{u})]\vert_{\ve{u}=0}\no\\
		&=\prod_\alpha [1+n_\alpha
		\partial_{u_\alpha}]\ii G^{K}(x,t;x',t'|\ve{u})]\vert_{\ve{u}=0}
	\end{align}
One can show from Eq.\ref{gku} that 
	\begin{equation}
	\ii\mathcal{G}^{K}(x,t;x',t')=\Gamma(x,x',t)-2\Lambda(x,x',t)
	\end{equation}
where we define the following quantities as
	\begin{align}
	\no\Lambda^\alpha(x,x',t)&= 
	-\frac{1}{2}\partial_{u_\alpha}[\ii G^K(x,t;x',t|\ve{u})]\vert_{u=0}=G^R(x,t;\alpha,0)[G^R(x',t;\alpha,0)]^\ast\\
	\Lambda(x,x',t)&= \sum_\alpha n_\alpha \Lambda^\alpha(x,x',t)\\
	\no \Gamma(x,x',t)&=\ii G^K(x,t;x',t|\ve{u}=0) = \sum_\alpha
	\Lambda^\alpha(x,x',t) + \ii\int_0^tdt_1\int_0^{t'}
	dt_2 G^R(x,t;x_1,t_1)\Sigma^K(x_1,t_1;x_2,t_2)G^A(x_2,t_2;x',t') 
	\end{align}
\end{widetext}
The initial condition specification $\{n_\alpha\}$ is entirely incorporated in $\Lambda$ alone, whereas $\Gamma$ is independent of the initial conditions and is fully determined by the system dynamics.

In the
next section, we will first integrate over the arguments of the Wigner
characteristic functions and then evaluate these multiple derivatives exactly to
present an analytic answer for R\'{e}nyi entropies of different orders (as
well as the Von-Neuman entanglement entropy) of subsystems of non-interacting open
quantum systems of Fermions. To our knowledge, this is the first time
such a general formula is being derived for open quantum systems.  An
alternative formulation, where the derivatives are first computed to
get the Wigner characteristic and the integrations are performed
afterwards, leads to a diagrammatic evaluation of entanglement
entropies. This will be shown in Appendix ~\ref{diag}.

\section{Exact Formula for Entanglement entropies of Open Quantum
  Systems\label{ExactFormula}}

In this section we will work out in detail the exact formula for the
second R\'{e}nyi entanglement entropy of an open quantum system. We will also provide
the final answers for the $n^{th}$ order R\'{e}nyi entropy, and hence for
the Von-Neumann entanglement entropy for the system. The detailed
derivation for this will be presented in Appendix ~\ref{app:Sn}.

The second R\'{e}nyi entropy of an open quantum system, starting from a
particular initial Fock state is given by
\begin{widetext}
  \begin{equation}
\mathrm{e}^{-S^{(2)}}= \prod_\alpha [1+n_\alpha
\partial_{u_\alpha}]\prod_\alpha [1+n_\beta \partial_{v_\beta}] 2^{V_A}\left[ \int {\cal D}[\args{\zeta}]{\cal
    D}[\args{\eta}]~ \mathrm{e}^{-\frac{1}{2}\sum^{'}
    (\bar{\zeta}_x,\bar{\eta_x})\left(\begin{array}{cc}
                                       \ii G^K(x,t;x',t|\ve{u}) & - \delta_{x,x'}%\hat{1}
                                       \\
                                     \delta_{x,x'}% \hat{1} 
                                     & \ii G^K(x,t;x',t|\ve{v}) \end{array}\right) \left(\begin{array}{c}
                                       \zeta_{x'}\\
                                                                                                        \eta_{x'} \end{array}\right)}\right]
 \end{equation}                                                                                               
 \end{widetext}
where $G^K(x,t;x',t|\ve{u})$ is given by Eq.~\ref{gku}. The gaussian
integrals can be performed easily to get
\begin{equation}
\mathrm{e}^{-S^{(2)}}=\prod_{\alpha}\left[1+n_{\alpha}\partial_{u_\alpha}\right]%
\prod_{\beta}\left[1+n_{\beta}\partial_{v_\beta}\right]%
\left.\left[\frac{1}{2^{V_A}}
\det\hspace{-2pt}\left[
\mathbb{M}
\right]
\right]\right|_{{\tiny\begin{array}{c}
		\ve{u}=0\\
		\ve{v}=0
		\end{array}}}
\label{eqnref:s2}
\end{equation}
where $\mathbb{M}$ is a $2V_A\times2V_A$ matrix defined as
\begin{equation}
\mathbb{M}\equiv%
\left(\begin{array}{cc}\mathbb{U}(\ve{u})&\mathbb{V}(\ve{v})\end{array}\right)%
:=\left(
\begin{array}{cc}
\ii \hat{G}^K(\ve{u}) &  -\hat{1}\\
\hat{1} & \ii \hat{G}^K(\ve{v}) 
\end{array}
\right)\label{eqref:S2mat}
\end{equation}
$\mathbb{U}$ \& $\mathbb{V}$ are $2V_A\times V_A$ matrices depending
on only $\ve{u}=\{u_\alpha\}$ and $\ve{v}=\{v_\beta\}$ respectively,
and we have suppressed the matrix indices of $G^K$ for notational
convenience. It is clear that
\begin{equation}
\mathbb{M}(\ve{u}=0,\ve{v}=0)=\left(
\begin{array}{cc}
\hat{\Gamma}&  -\hat{1}\\
\hat{1} & \hphantom{-}\hat{\Gamma} 
\end{array}
\right)
\end{equation}
Note that the $u_\alpha$ derivatives only act on $\mathbb{U}$ and
similarly $\{\partial_{v_\beta}\}$ only act on $\mathbb{V}$. Hence we
can treat them separately and focus
solely on the action of $\prod_\alpha [1+n_\alpha \partial_{u_\alpha}]
$ on $\det\mathbb{M}$ for the sake of illustration.

\begin{widetext}

\begin{figure}[h!]
	\centering
	\includegraphics[width=\textwidth]{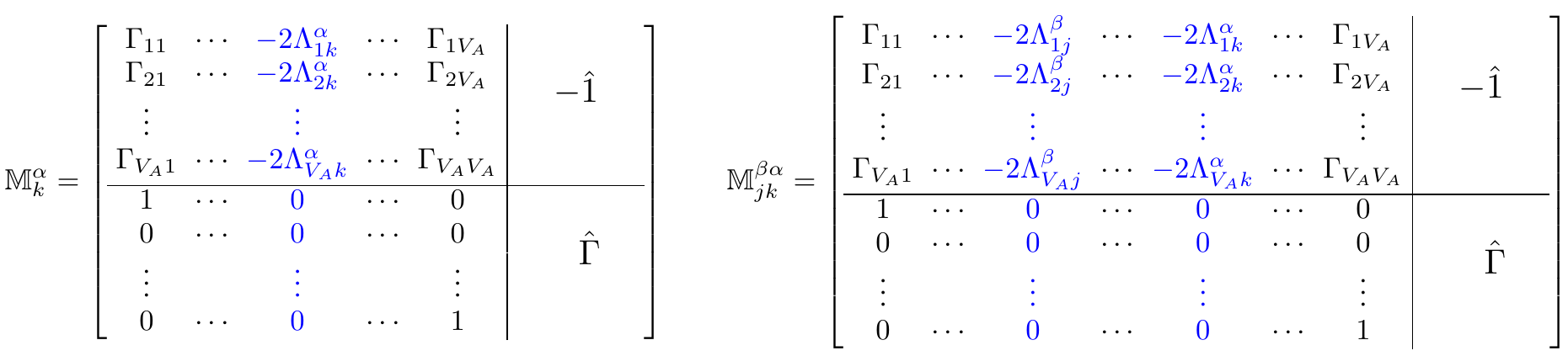}
	\caption{The matrix M obtained by taking u-derivatives of the determinant in Eq. 30 Note that the full answer for R\'{e}nyi entropy is obtained by taking a sum of determinants of many such matrices.  See text for details.}
	\label{Matstruct}
\end{figure}

\end{widetext}
The first thing to note is that $\partial_{u_\alpha}\partial_{u_\beta}
G^K(x,t;x',t|\ve{u})=0$, i.e. there can only be a single derivative of a
particular matrix element. Let us consider a single $u_\alpha$ derivative
acting on $\text{det} ~\mathbb{M}$. Consider a matrix where the matrix
elements in all columns except the $x^{th}$ one ($x\leq V_A$) are same
as $\mathbb{M}$, while the matrix elements in the $x^{th}$ column are
replaced by their $u_\alpha$ derivatives. Let us call this matrix
$\mathbb{M}^\alpha_x$. If we set $\ve{u}=\ve{v}=0$ in this matrix, we
will replace the $x^{th}$ column by $\{-2\Lambda^\alpha(i,x)\}|_{i=1}^{V_A}$ on the top
half and a $0$ vector in the bottom half. This matrix is shown in
Fig~\ref{Matstruct}. It is then easy to show that the derivative of
the determinant is the sum of the determinant of these matrices, from
$x=1$ to $x=V_A$
\beq
\partial_{u_\alpha} |\mathbb{M}| =\sum_{x=1}^{V_A} |\mnm^\alpha_x|
\eeq
Here, for brevity we denote $|\ast|:=\det[\,\ast\,]$. One can take this argument forward to show that
\beq
\partial_{u_\alpha}\partial_{u_\beta} |\mathbb{M}| =\sum_{x_1\neq x_2}|\mnm^{\beta\alpha}_{x_1x_2}|
\eeq
where $\mnm^{\beta\alpha}_{xx'}$ is the matrix $\mnm$ with the $x_1^{th}$
column replaced by its $u_\beta$ derivative and the $x_2^{th}$ column
replaced by its $u_\alpha$ derivative. For $(\ve{u},\ve{v})=0$, this
matrix is shown in Fig~\ref{Matstruct}. The $x_1^{th}$ column is
replaced by $-2\Lambda_\beta(k,x_1)$ and the $x_2^{th}$ column is replaced
by  $-2\Lambda^\alpha(k,x_2)$. One can extend this construction for
higher order derivatives to get 
\begin{widetext}
\begin{equation}
\begin{aligned}
&\prod_{\alpha}\left[1+n_{\alpha}\partial_{u_\alpha}\right]\lvert \mathbb{M}\rvert=\left[1+\sum_{\alpha}n_{\alpha}\partial_{u_\alpha}+\frac{1}{2!}\sum_{\alpha\neq\alpha'}n_{\alpha}n_{\alpha'}\partial_{u_\alpha}\partial_{u_{\alpha'}}+\dotsb\right]\lvert \mathbb{M}\rvert\\
&=\lvert \mathbb{M}(0)\rvert%
+\sum_{x=1}^{V_A}\sum_{\alpha=1}^{N}\lvert \mathbb{M}^{\alpha}_{x}\rvert%
+\frac{1}{2!}\sum_{x_1\neq x_2}\sum_{\alpha_1\neq\alpha_2}\lvert
\mathbb{M}^{\alpha_1\alpha_2}_{x_1x_2}\rvert+\dotsb+\frac{1}{V_A!}\sum_{\substack{x_1\ldots x_{V_A}\\x_i\neq x_j}}\sum_{\substack{\alpha_1\ldots \alpha_{V_A}\\\alpha_i\neq\alpha_j}} \lvert \mathbb{M}^{\alpha_1\ldots\alpha_{V_A}}_{x_1\ldots x_{V_A}}\rvert.%
\end{aligned}\label{eqnref:seriesU_con}
\end{equation}
where each of the terms above are evaluated at $\ve{u}=0$. At this
point we note that $\hat{\Lambda}_\alpha$ has a factorizable form,
i.e. $\Lambda^\alpha(x,x',t) \sim g_\alpha(x)
g^\ast_\alpha(x')$. Hence, if any of the $\alpha$ indices are
repeated in a matrix of the form
$\mnm^{\alpha_1\alpha_2...}_{x_1,x_2..}$, the corresponding columns
are proportional to each other. As a result, the determinant of such a
matrix is $0$. This allows us to replace the constrained sums on
$\alpha$ indices in
Eq.~\ref{eqnref:seriesU_con} by unconstrained sums. The added terms
are zero, since they involve identification of $\alpha$ indices. The
sums over $\alpha$ indices can be done and we can replace
$\hat{\Lambda}^\alpha$ by $\hat{\Lambda}$ in each of the
matrices. This defines a new matrix $\mnm_{x_1..x_k}$ where the $x_1$,
... $x_k$ columns of $\mnm$ are replaced by the corresponding columns
of $\mathbb{L}=\begin{pmatrix} -2\hat{\Lambda} &\hat{0}\\ \hphantom{-}\hat{0} &\hat{0} \end{pmatrix}$. Using this, we finally get
\begin{equation}
\begin{aligned}
\prod_{\alpha}\left[1+n_{\alpha}\partial_{u_\alpha}\right]\lvert \mathbb{M}\rvert|_{\ve{u}=\ve{v}=0}&=%
\lvert \mathbb{M}(0)\rvert%
+\sum_{x=1}^{V_A}\lvert \mathbb{M}_{x}\rvert%
+\frac{1}{2!}\sum_{x_1\neq x_2}\lvert \mathbb{M}_{x_1x_2}\rvert+\ldots+\frac{1}{V_A!}\sum_{x_1\ldots x_{V_A}} \lvert \mathbb{M}_{x_1\ldots x_{V_A}}\rvert=\lvert \mathbb{M}(0)+\mathbb{L}\rvert%
\end{aligned}
\end{equation}\label{eqref:reconstitute}
\end{widetext}
Note that in the last line, sum of  $2^{V_A}$ determinants have been
reconstituted a single determinant. The easiest way to see this is to
expand a determinant of sum of two matrices in terms of the matrix
elements and regroup the terms. A similar procedure follows for the
$\{v_\beta\}$ derivatives acting on $\mathbb{V}$, with columns of
$\mnm$ being replaced by that of
$\bar{\mathbb{L}}=\left(\begin{array}{cc} \hat{0} &\hphantom{-}\hat{0}\\ \hat{0}
                                            &-2\hat{\Lambda} \end{array}\right)$.
Putting these results together into eqn.\eqref{eqnref:s2}, we have
\begin{equation}
\begin{aligned}
\mathrm{e}^{-S^{(2)}}=\frac{1}{2^{V_A}}
\det\hspace{-2pt}\left[
\begin{array}{cc}
\hat{\Gamma}-2\hat{\Lambda}& -\hat{1}\\
\hat{1} & \hat{\Gamma}-2\hat{\Lambda}\\
\end{array}
\right]=
\det\left[\frac{\hat{1}+(\hat{\Gamma}-2\hat{\Lambda})^2}{2}\right]
%\det\left[\frac{\hat{1}+(\ii\hat{\mathcal{G}}^K)^2}{2}\right]
\end{aligned}
\end{equation}
and hence
\begin{equation}
\begin{aligned}
S^{(2)}&=-\tr{\ln\left[\left(\frac{\hat{1}-\hat{\Gamma}}{2}+\hat{\Lambda}\right)^2+%
    \left(\frac{\hat{1}+\hat{\Gamma}}{2}-\hat{\Lambda}\right)^2\right]}
\end{aligned}
\label{S2_final}
\end{equation}
The derivation can be readily extended for
higher integer R\'{e}nyi indices; for $n\geq2$, we get
	\begin{equation}
	S^{(n)}=\frac{1}{1-n}\tr{\ln\left[\left(\frac{\hat{1}-\hat{\Gamma}}{2}+\hat{\Lambda}\right)^n+%
		\left(\frac{\hat{1}+\hat{\Gamma}}{2}-\hat{\Lambda}\right)^n\right]}\label{eqnref:Sn}
	\end{equation}
Details about the derivation of the above are presented in appendix
\ref{app:Sn}. We can then analytically continue the expression for
$S^{(n)}$ in eqn.\eqref{eqnref:Sn} to take the $n\to1$ limit and
recover the von-Neumann entanglement entropy,
\begin{widetext}
	\begin{equation}
	S_{\text{vN}}=-\tr{%
	\left(\frac{\hat{1}-\hat{\Gamma}}{2}+\hat{\Lambda}\right)\ln\left(\frac{\hat{1}-\hat{\Gamma}}{2}+\hat{\Lambda}\right)+%
	\left(\frac{\hat{1}+\hat{\Gamma}}{2}-\hat{\Lambda}\right)\ln\left(\frac{\hat{1}+\hat{\Gamma}}{2}-\hat{\Lambda}\right)%
	}. \label{SVN_final}
	\end{equation}
      \end{widetext}
These are exact results for the time dependent evolution of the
entanglement von-Neumann and R\'{e}nyi entropies of a generic gaussian fermionic open quantum system
initialized to an arbitrary Fock state. They are the key new results
presented in this paper. We note that these results are strictly valid
when the number of particles $N$ is larger than the subsystem size
$V_A$. In this case the reduced density matrix has support in the whole
Fock space of the degrees of freedom in the subregion $A$, whereas
when $N<V_A$, some states in the Fock space cannot be accessed. In the
thermodynamic limit, $V\rightarrow \infty$ and $N\rightarrow \infty$,
with a fixed density $N/V=\rho$. So, if $V_A\rightarrow \infty$ in a
way that $V_A/V \rightarrow 0$ (this is the limit in which answers fro
conformal field theories hold), the results above are always valid. If
$V_A/V \rightarrow f$, i.e. the subsystem is a finite fraction of the
system size, the results will continue to hold when $\rho > f$.

\section{Entanglement Entropy of Fock states\label{Fock}}

In this section, we will focus our attention on the
entanglement entropy of a class of Fermionic Fock states. In our
formalism, this can be obtained by calculating the entanglement entropy
at $t=0$. In absence of the dynamics, there is no
difference between open and closed quantum systems.
% and our calculations recover well known formulae based on correlation matrix approach~\cite{Peschel_2003}.

% 
\begin{figure}[t!]
	\includegraphics[width =\columnwidth]{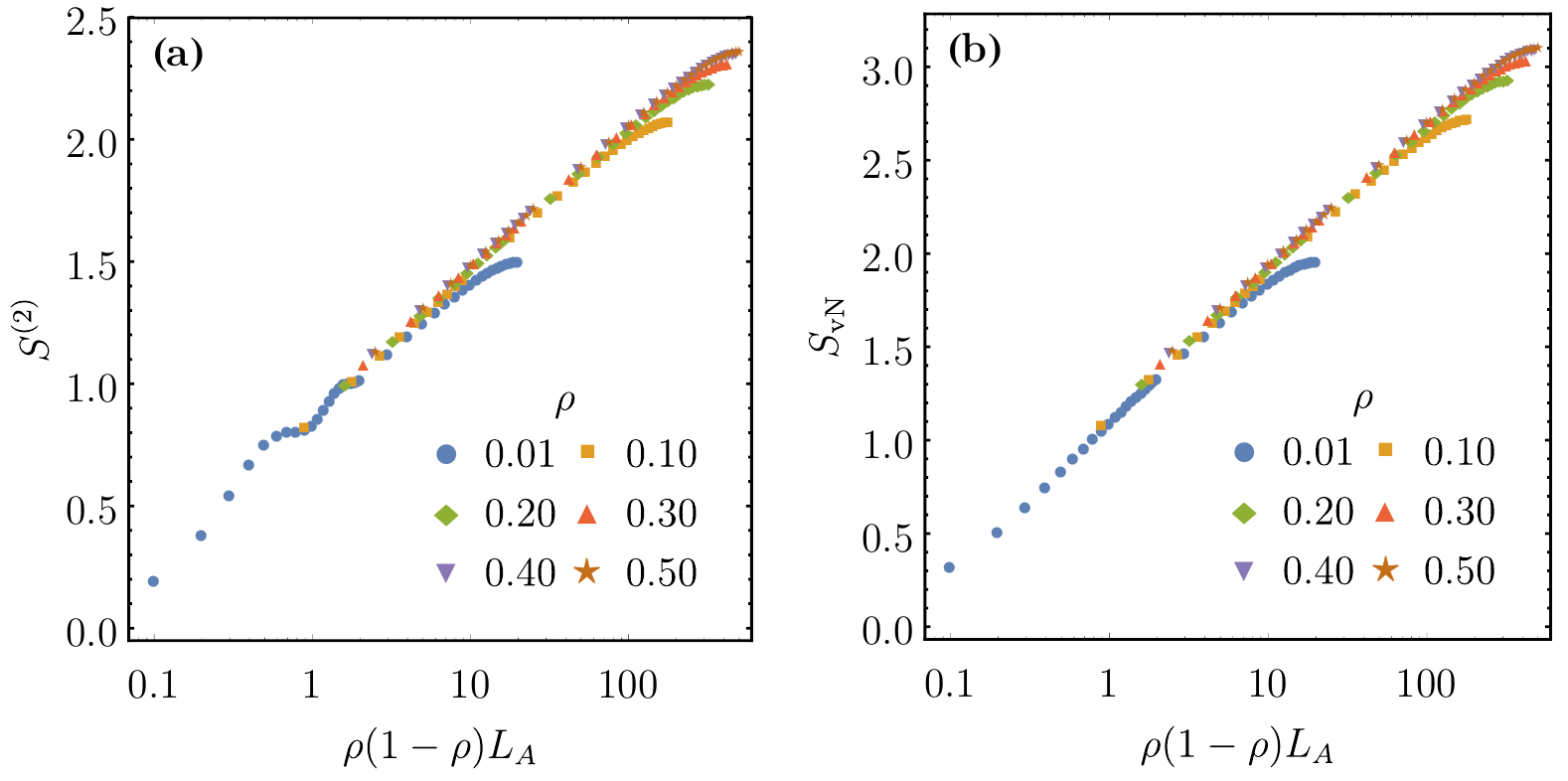}
	\caption{ The scaling of (a) second R\'{e}nyi entropy and (b) the
		von-Neumann entanglement entropy with the subsystem size for a
		Fermi sea of one dimensional spinless Fermions. Note the
		logarithmic scaling with the system size. The entanglement entropy
		vs subsystem size graph for Fermions at different densities $\rho$
		collapse to a single graph when plotted as a function of
		$\rho(1-\rho)L_A$, where $L_A$ is the size of the subsystem}
	\label{Fig:FS_Ent_Scaling}
\end{figure}

For a closed non-interacting system, $G^R(x,t;\alpha,0) =-i\Theta(t)
\sum_{\mu} \phi_\mu(x)\phi^\ast_\mu(\alpha) \mathrm{e}^{-iE_\mu t}$, where $\mu$
denotes the eigenstates of the single-particle Hamiltonian with
wavefunction $\phi_\mu$ and energy $E_\mu$. This leads to 
\beq
\Gamma(x,x',t)= \sum_\alpha G^R(x,t;\alpha,0)\left[G^R(x',t;\alpha,0)\right]^\ast=\Theta(t)\delta_{x,x'},
\eeq
where we have used the orthonormality of wavefunctions $\sum_\alpha
\phi_\mu(\alpha) \phi^\ast_\nu(\alpha)=\delta_{\mu,\nu}$ to get this
answer.  In this case, the entanglement entropies are given by
%\begin{widetext}
  \bqa
 \no S^{(n)}&=& \frac{1}{1-n}\tr{\ln\left[\hat{\Lambda}^n+( \hat{1}-\hat{\Lambda})^n\right]}\\
S_{vN}&=&- \tr{\left [ \hat{\Lambda} \ln \hat{\Lambda} +
  (1-\hat{\Lambda})\ln (1-\hat{\Lambda})\right]}
 \eqa
%\end{widetext}
% 
\begin{figure}[t]
	\includegraphics[width =\columnwidth]{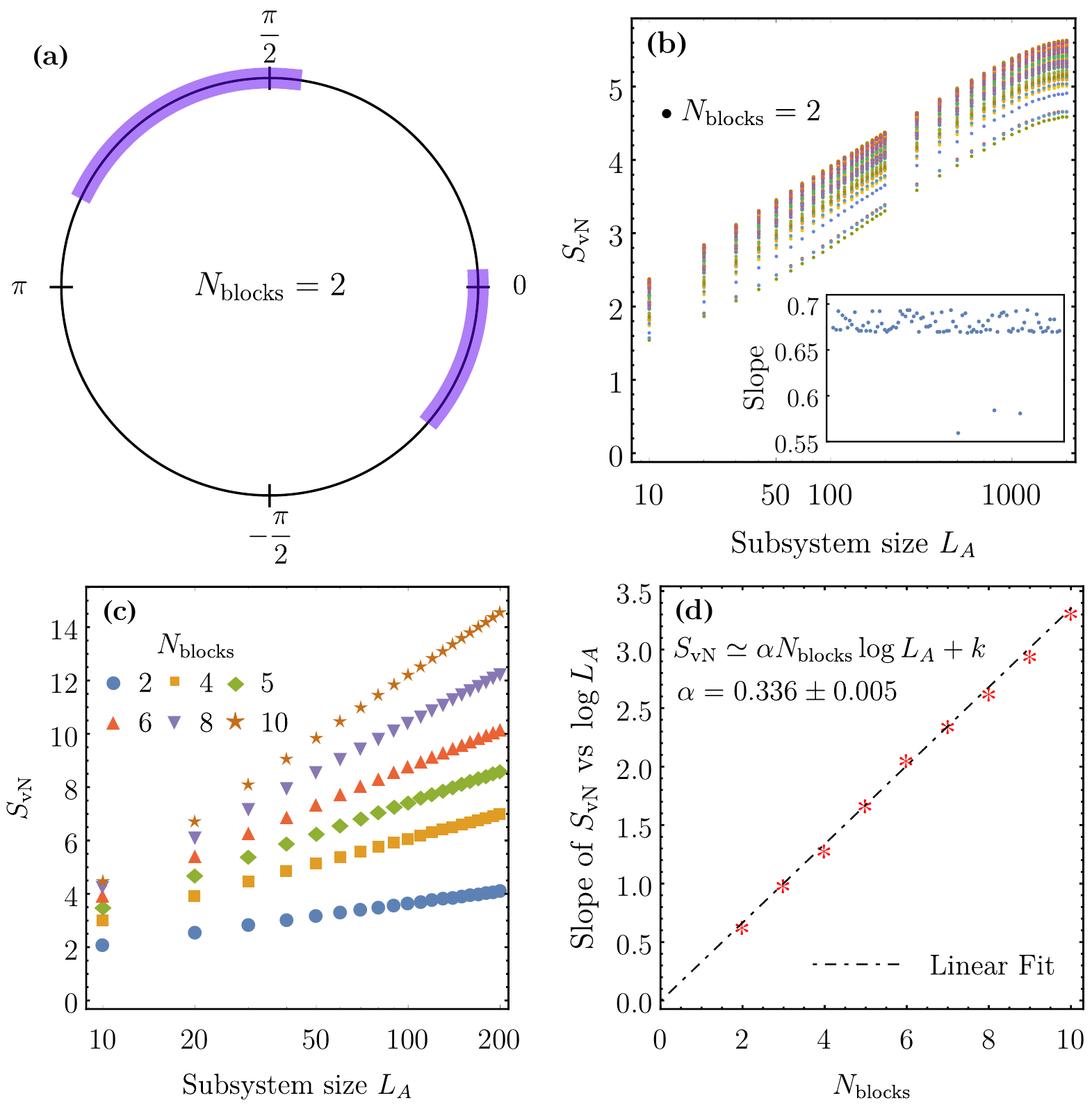}
	\caption{ (a) A momentum Fock state of spinless Fermions in one
		dimension with $2$ contiguous blocks of occupied momenta. Note
		that under periodic boundary conditions, the Brillouin zone is
		mapped to a circle with circumference $2\pi/a$, where $a$ is the
		lattice spacing. (b) The scaling of von-Neumann entanglement
		entropy with subsystem size for Fock states with $2$ contiguous
		blocks. The slope of the logarithmic scaling is twice that for the
		Fermi sea ($y_2=2\times c/3$), where $c=1$ is the central charge
		for free Fermions. Inset: the value of the slope for each Fock state with $2$
		contiguous blocks (plotted with configuration number). Note the lack
		of scatter in the slope, showing each $2$-block state has the same
		slope independent of the size or position of the blocks.(c) The
		scaling of entanglement entropy with subsystem size for Fock states
		with $p$ contiguous blocks with $p=2, 4, 5, 6, 8, 10$. These states show
		logarithmic scaling with a coefficient which increases with $p$. (d)
		The coefficient of the logarithmic scaling plotted as a function of
		the number of blocks $p$, showing that the coefficient is simply
		$y_p=p \times c/3$. All the above data is for asystem with $L=4096$
		sites and a density of $\rho=0.3$}
	\label{Fig:BlockStates_Ent_Scaling}
\end{figure}
While the above formulae are true for any closed system dynamics starting
from Fock states, we will focus at $t=0$ where $\Lambda(x,x',t=0)= \sum_\alpha
n_\alpha^0 \phi_\alpha(x)\phi^\ast_\alpha(x')$. Here $n_\alpha^0$ is the
occupation number of the mode $\alpha$  in the initial Fock state, and one gets
the formula derived by Peschel et. al ~\cite{Peschel_2003} using
correlation matrix approach.

\begin{figure}[t!]
	\includegraphics[width=\columnwidth]{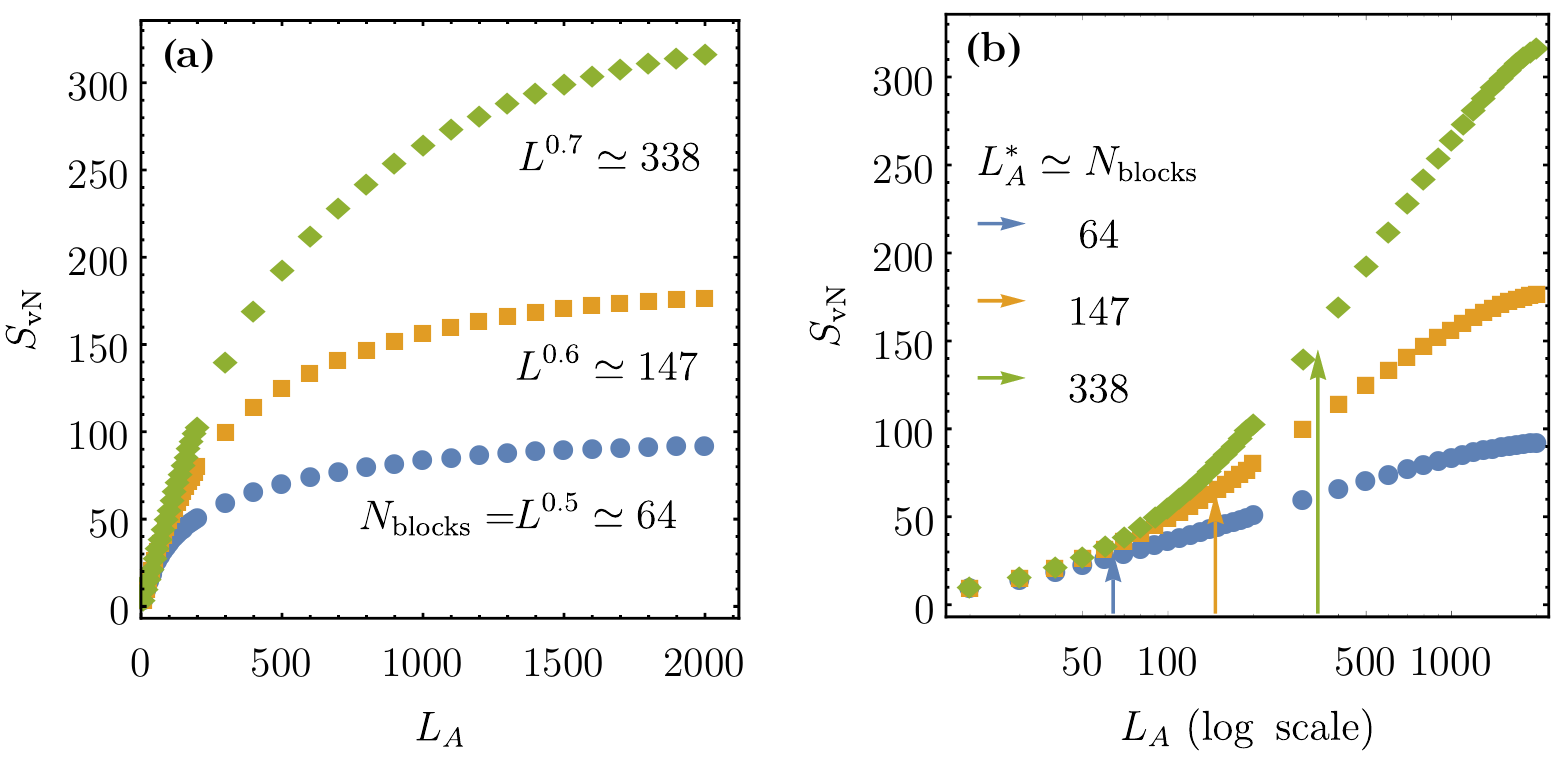}
	\caption{The scaling of entanglement entropy of momentum Fock states
		with subsystem size $L_A$ for states with $N_b=64$, $147$ and $338$ occupied
		blocks in a system with $4096$ lattice sites and a density of
		$\rho=0.3$. In (a) the subsystem size is on a linear scale, while in
		(b) the subsystem size is on a logarithmic scale. Note that at small
		values of $L_A$, the size dependence is linear (see (a)), which
		transitions to a logarithmic scaling for $L_A >N_b$, as seen in
		(b). The same states show critical or linear behaviour depending on
		the subsystem size.}
	\label{Fig:EEScaling_log_linear}
\end{figure}
We would like to point out one thing at the outset. If we consider a
set of states $\{|n\rangle\}$, calculate the R\'{e}nyi/von Neumann entanglement entropy for each
one (with the same subsystem), and sample the entropy of the state
$|n\rangle$ with probability $p_n$, this is not equal to the
entanglement entropy of the density matrix
$\hat{\rho}=\sum_{\{|n\rangle\}}p_n |n\rangle \langle n|$. For
example, if
$\hat{\rho}^{(n)}_A$ is the reduced density matrix obtained from
$|n\rangle$, the first case yields $\mathrm{e}^{-S^{(2)}}=\sum_n p_n Tr
\left[\hat{\rho}^{(n)}_A\right]^2$, while the second case yields $\mathrm{e}^{-S^{(2)}}= Tr
\left[\sum_n p_n\hat{\rho}^{(n)}_A\right]^2$. This is in contrast to
correlation functions, where these two procedures will yield the same
result. {\it For example, sampling all states with equal probability will
not be equivalent to calculating entanglement entropy for an infinite
temperature ensemble.} 

We consider momentum Fock states in a one dimensional
lattice of spinless Fermions with periodic boundary conditions.  In
the thermodynamic limit, the
momentum states are defined on a  circle
of circumference $2\pi/a$, where $a$ is the lattice spacing. In this
case, $\hat{\Lambda}$ is a Toeplitz matrix
generated by the momentum distribution (i.e.  $\Lambda(x,x',t=0)$ is the Fourier
transform of the momentum distribution ).
%
%$Tr ~\hat{\rho}_A^2$ is the product of determinants of two
%matrices, (
%$\hat{1}\pm \ii [\hat{1}-2\hat{\Lambda}]$), each of which
%is a Toeplitz matrix.
%
One can then use the Fisher Hartwig conjecture~\cite{Basor1994_GeneralizedFisherHartwig,Alba_2009},
which relates the jump discontinuities in the generating function of a
Toeplitz matrix to a power law scaling of its determinant with the
dimension of the matrix. In this case, it relates the jump
discontinuities to
a logarithmic scaling of the entanglement entropies with the subsystem
size $L_A$ (which is the dimension of the matrix $\Lambda$). For example, it is well known that the Fermi sea , which is a
contiguous line of momentum occupancies with two jump discontinuities
at the two ends, leads to a logarithmic dependence of the entanglement
entropy with the subsystem size, with a coefficient related to the
central charge of the corresponding conformal field theory, $c$: $S^{(2)}=
\frac{c}{4} \ln~ L_A$ and $S_{vN} = \frac{c}{3} \ln~
L_A$~\cite{Calabrese_2004}, where $c=1$ for two species (left and
right moving) of fermions in the system.

We consider a system of $L=4096$ lattice sites with different
number of particles $N$, leading to different
densities $\rho= N/L$. In Fig ~\ref{Fig:FS_Ent_Scaling}(a) and (b), we
plot the R\'{e}nyi and von
Neumann entanglement entropy of the one dimensional Fermi sea as a
function of the subsystem size $L_A$ to show the logarithmic
scaling. We note that the curves for different densities collapse
when plotted as a function of $\rho(1-\rho) L_A$. For closed fermionic
systems, one can either describe the system in terms of particles created on top of a
vacuum state with zero particles, or in terms of holes created on top
of a state with all single particle modes filled. Thus, there is an
invariance under $\rho \rightarrow 1-\rho$, which is reflected in the
collapse of the curves when plotted as a function of
$\rho(1-\rho)L_A$.

We now consider a Fock state made of two contiguous blocks of occupied
momentum states. To see what this means, we show a Fock state with $N_b=2$
contiguous blocks of occupation in Fig~\ref{Fig:BlockStates_Ent_Scaling}(a). The entanglement entropy
of this state is $S_{vN} =2\times \frac{c}{3} \ln~
L_A$, as there are now four jump discontinuities in the momentum
distribution. This answer does not depend on the size of the occupied
blocks, their locations or the separation between them, and only cares about the number of jump discontinuities, just as Fisher
Hartwig conjecture would predict. In Fig.~\ref{Fig:BlockStates_Ent_Scaling}(b), we plot the
entanglement entropy of several of these ``2-block'' states with the subsystem
size. The logarithmic scaling is obtained with a coefficient which is twice
the coefficient for the Fermi sea. The value of this coefficient (slope of the
curve), obtained for different ``2-block'' states, is shown in the
inset of Fig.~\ref{Fig:BlockStates_Ent_Scaling}(b) as a function of
the configuration number for about $100$ such states. The absence of
scatter shows that each state follows the logarithmic scaling with the
same co-efficient. We note that the intercept of the curve is
non-universal and varies widely from one Fock state to another.

This argument can now be extended to the case
of Fock states with $p$ contiguous blocks of momentum occupancy, which
will have  $S_{vN} =p\times \frac{c}{3} \ln~
L_A$. In Fig.~\ref{Fig:BlockStates_Ent_Scaling}(c), we plot the entanglement entropy of Fock
states with $p$ contiguous blocks of occupancy for $p=2,4, ..10$ as a
function of $L_A$ and show that the logarithmic scaling with the
subsystem size is recovered. In Fig.~\ref{Fig:BlockStates_Ent_Scaling}(d), we plot the
slope of the curve with $p$ to see that the coefficient matches our
expectations. For each value of $p$ we have
averaged over  $ 20$ random configurations of the blocks and the variation of the slope from configuration to configuration is negligible.
% The subleading corrections (the intercept of the curve) however fluctuates a lot from sample to sample. 
This suggests a classification of the ``critical
states'' in terms of number of contiguous occupied blocks in the
momentum space, $N_{b}$. We note that this classification is similar to the one proposed by
Ref.~\onlinecite{Alba_2009} and is
different from that of Jafarizadeh {\it et. al}~\cite{Rajabpour_excitedstates}, since
our states do not have any periodic arrangements in momentum space.

However, when we extend the above construction to states with larger
number of contiguous blocks, a more comprehensive picture emerges. If
we consider the entanglement entropy of Fock states with a particular
value of $N_{b}$ as a function of the subsystem size $L_A$ (for
$L_A/L \ll 1$), then:{\it  (i) $S_{vN}(N_{b}) \sim N_{b} \times \frac{c}{3} \ln~
L_A$  for $L_A \gg N_{b}$ and (ii) $S_{vN}(N_{b}) \sim L_A$
for $L_A \ll N_{b}$, and the dependence on the subsystem size has
a smooth crossover from linear to logarithmic around $N_{b} \sim
L_A$. } This is shown in Fig~\ref{Fig:EEScaling_log_linear}(a) and (b), where the
entanglement entropy of states with $N_{b}=64$, $147$ and $338$
are plotted as a function of $L_A$ on linear and logarithmic scales
respectively. We have used a system size of $L=4096$ and a density of
$\rho=0.3$ for these plots. We clearly see that the dependence goes
from linear to logarithmic and the change happens at $L_A \sim
N_{b}$.  We note that the Fisher Hartwig conjecture works for a
finite number of singularities (compared to size of the matrix), and
hence it is not surprising that it breaks down when number of
discontinuities in the generating function is larger than the system
size.

The key insight that we get from the above exercise is that there are
no ``critical'' or ``non-critical'' states. The same quantum state can
show either critical (logarithmic dependence on $L_A$) or non critical
(linear dependence on $L_A$) behaviour of entanglement entropy
depending on the size of the subsystem. For a given subsystem size,
however, we can divide states into those which show logarithmic and
linear dependence of entanglement entropy. Every
Fock state can be written as the ground state of some non-interacting
Hamiltonian and Alba et. al~\cite{Alba_2009} had showed that if this Hamiltonian is
long ranged, then the entanglement entropy scales linearly with
subsystem size, while a short range Hamiltonian results in a
logarithmic scaling. In that language, the above results show that if the subsystem size is longer
than the range of this Hamiltonian, the state will show critical
behaviour in the scaling of the entanglement entropy. Note that this
also points out the difficulty of obtaining these estimates
numerically, since there is a strong subsystem size dependence of the results.

We will
now estimate the number of these ``critical''  states (for a given
$L_A$, and states with $N_b < L_A$) in a
system with $N$ particles on $L$ sites. The first thing to note is
that in general $1 \leq N_b \leq \min(N,L-N)$. Since each occupied
block has to be followed by an empty block, the number of blocks is
bounded by the number of particles/holes in the system. Let us first
consider the number of possible states with $N_b$ blocks. This
problem is equivalent to the number of ways of choosing $N_b$ out of
$N$ positions amongst the occupied momenta to place the boundaries of the blocks, multiplied by the number of ways of choosing $N_b$ out of
$L-N$ positions amongst the unoccupied momenta to assign the gaps between the occupied blocks. So
the total number of states with $N_b=Lx$ blocks is
\bqa
\Omega(N_b)& =&\frac{N!}{(N-N_b)!N_b!} \frac{L-N!}{(L-N-N_b)!N_b!}\\
\nonumber &\sim & \frac{\mathrm{e}^{-L[2x \ln x +(\rho-x) \ln (\rho-x)
    +(1-\rho-x)\ln (1-\rho-x)]}}{\mathrm{e}^{-L[\rho \ln \rho
    +(1-\rho)\ln (1-\rho)]}}
\eqa
where we have used the thermodynamic limit with $N=L\rho$. in the last
line. The number of critical states is obtained by summing over this
expression for $0<x< \alpha$, where the subsystem size
$L_A=L\alpha$. Now in general one requires $ \alpha \ll 1$ for the
critical logarithmic scaling to hold (this is the limit in which
conformal invariance is preserved). So we expect an exponentially
large number of ``critical'' states, although they will form a
vanishing fraction of the total number of states, unless $L_A/L$ is a
substantial fraction, in which case one needs to worry about
corrections due to finite $L_A/L$.

\section{Entanglement Dynamics in Open Quantum Systems\label{OQS}}

In this section, we will finally use our formalism to compute the
dynamics of entanglement entropy in an open quantum system of
Fermions. We will see how the entanglement entropy bears the
signatures of different classical and quantum processes during the
evolution of the system.

We consider a one dimensional lattice of spinless Fermions with
nearest neighbour hopping and free boundary conditions
\beq
H_s= -g\sum_{i=1}^{L-1} c^\dagger_{i+1}c_i + h.c.
\eeq
where $i$ indicate the lattice site and $g$ is the hopping amplitude which
sets the bandwidth of the system. At $t=0$, the system is in a Fock
state described by occupation number of each lattice site. The particular
initial condition we choose here is the following: Sites $1$ to $L/2$
are filled with $1$ particle, while sites $L/2+1$ to $L$ are
empty. This is shown in Fig.~\ref{Fig:Neqbm_density}(b).  We
note that the spinless Fermi gas in one dimension can be mapped to a
spin system. In that language, our initial condition corresponds to a domain
wall at the center of the system.

\begin{figure}[t!]
	\includegraphics[width=\columnwidth]{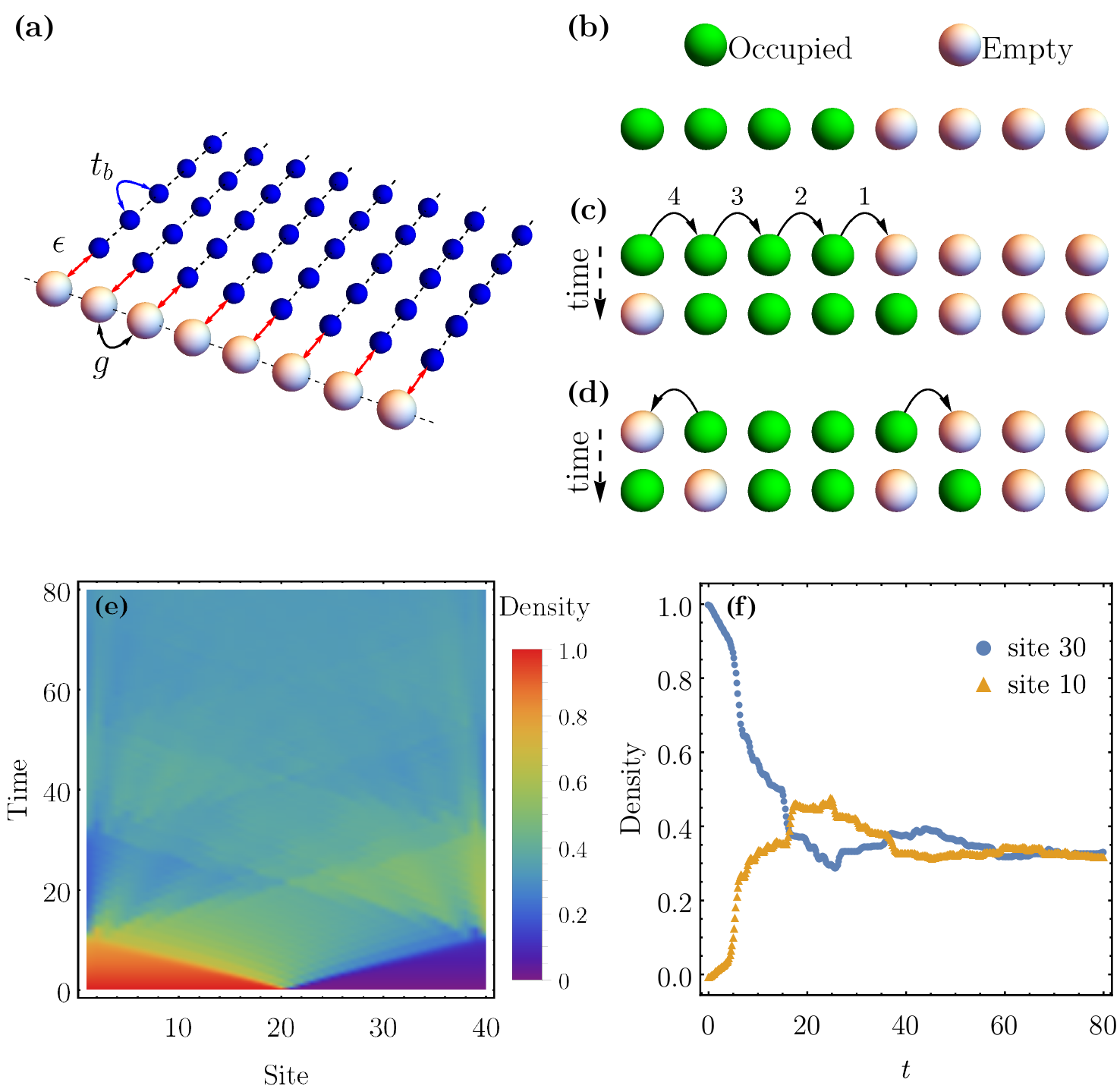}
	\caption{ (a) A schematic diagram of a one dimensional Fermionic chain with nearest neighbour hopping $g$, with each site coupled to a Fermionic bath represented by one dimensional chains with hopping $t_B$. The system bath coupling is $\epsilon$.
		(b) The initial real space Fock state with the left half of
		the system filled with $1$ particle per site and the right half
		kept empty. The system has $L=40$ sites. There is a domain wall at the center of the
		system. (c-d) The different physical processes during the
		non-equilibrium dynamics of this open quantum system: (c) Coherent motion of
		a domain wall wavefront across the system and (d) Breaking of a single domain
		into multiple domains and vice-versa. In addition every site can exchange energy and particles with the local bath (e) A color plot of the evolution of the
		density profile of the system after it is coupled to an external bath
		at $t=0$. The propagation of the wavefront of
		domain walls  lead to the characteristic diamond shapes in the
		color plot. (f) The density profile of two sites located
		symmetrically from the center on the left and right. The initial
		in(de)crease is due to exchange of particles with the bath,
		whereas the sharp
		jumps correspond to the passing of a domain wavefront. The
		oscillations correspond to breaking up of a single domain into two
		and vice-versa. The coherent motion gets damped and settles into a
		homogeneous value for the density at long times. The graphs
		correspond to parameters $g=1.0$,
		$t_B=2.0$, $\epsilon=0.2$. The bath
		temperature is $T=1.0$ and chemical potential $\mu=-1.11$, so that the
		equilibrium density of the system is $\rho_{eq}=0.32$ in this case.}
	\label{Fig:Neqbm_density}
\end{figure}

At $t=0$ we also turn on the coupling of this system to an external
bath of Fermions with which it can exchange energy and particles. Each site of the system couples to a bath, which is
modelled as another linear chain of free Fermions with a nearest
neighbour hopping scale
$t_B$. The system bath coupling is linear and is controlled by a
parameter $\epsilon$. A schematic of the system , the bath and the system bath coupling is shown in Fig 5(a). The details of this model is the same as that
used in Ref~\onlinecite{Ahana_OQS}, which was used to study the dynamics of
correlation functions in the system. The effect of the bath on our
system is characterized by the spectral function of the bath
$J(\omega)$, its temperature $T$ and its chemical potential $\mu$. In
the long time limit, we expect our system to thermalize with this bath
with a density determined by $T$, $\mu$ and $g$. For our specific
model of the bath, the spectral density is given by $J(\omega)=
\Theta(4t_B^2-\omega^2)\frac{2}{t_B}\sqrt{1-\omega^2/4t_B^2}$, with a
band width of $4t_B$. The band edge singularities of this spectral
function leads to non-Markovian dynamics in this system
~\cite{Ahana_OQS}. Throughout this section, we will consider a
lattice of $L=40$ sites. Our system will be characterized by a hopping strength $g=1.0$, while the bath
is characterized by a hopping $t_B=2.0$ (making sure bath bandwidth is
larger than system bandwidth, so that it acts as a heat bath for all
modes in the system), a temperature $T=1.0$ ( we take $k_B=1$) and a chemical potential
$\mu=-1.1$, which corresponds to a bath particle density of $0.4$. If
the system thermalizes with the bath, it should have a density of
$0.32$, as compared to an initial density of $n=0.5$. The system bath
coupling is set to $\epsilon=0.2$. 

\begin{figure}[t!]
	\includegraphics[width=\columnwidth]{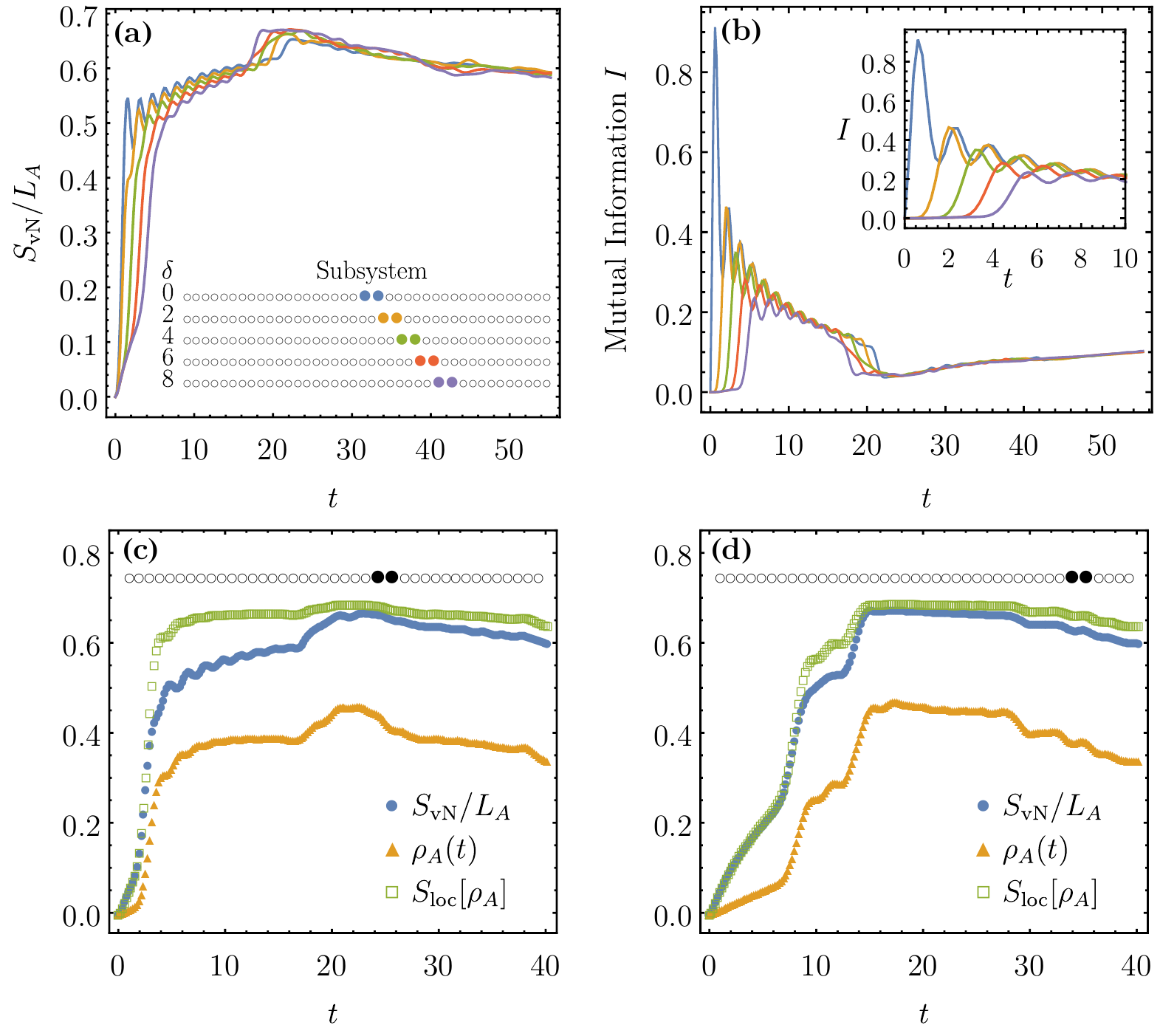}
	\caption{(a) The time evolution of entanglement entropy of $2$-site
		subsystems located at various distances $\delta$ from the center
		of the system. The subsystems are shown as colored dots, and the
		entropy of the corresponding subsystem is plotted with same color.
		After an initial common rise, the curves show a sudden rise at
		different times corresponding to the time the wave front passes
		through this subsystem. (b) The mutual information between the two
		$1$-site subsystems that make up the $2$ site subsystems shown in
		(a). The sharp rise and the oscillations are prominent
		here. Inset: a close up of the small time dynamics of the mutual
		information showing that it rises at different times for
		subsystems at different distances from the center. (c) and (d) The
		time evolution of average density $\rho_A$ (yellow) and entanglement
		entropy density $S_{vN}/L_A$ (blue) of a $2$-site subsystem starting at the (c)
		$24^{th}$ and (d)$35^{th}$ site. The green curve is $S_{loc}$,
		which is the entropy of an effective one site subsystem with the
		average density. The graphs
		correspond to a system hopping $g=1.0$, a bath hopping
		$t_B=2.0$, a system bath coupling $\epsilon=0.2$. The bath
		temperature $T=1.0$ and chemical potential $\mu=-1.11$, so that the
		equilibrium density of the system is $\rho_{eq}=0.32$ in this case.}
	\label{Fig:Neqbm_Ent_2site}
\end{figure}

\begin{figure}[t!]
  \includegraphics[width=\columnwidth]{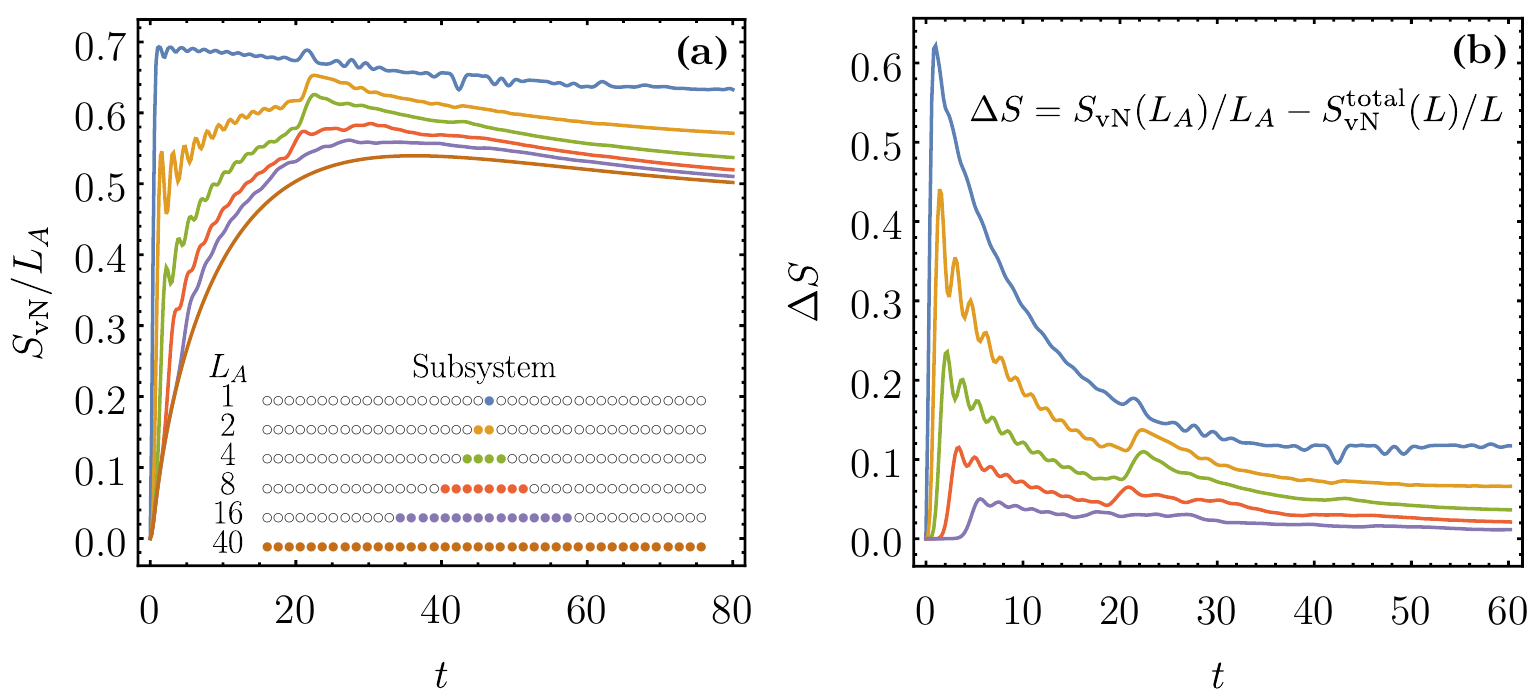}
  \caption{(a) Evolution of entanglement entropy (per site) $S_{vN}/L_A$ of
    subsystems, of different sizes, $L_A=1$, $2$, $4$, $8$, $16$, and
    $40$ (whole system), located in the
    central part of the system. Note the sharp jump occuring at
    different initial times for different subsystems coupled with
    oscillations. These features are absent when $L_A=40$, i.e. the
    whole system is considered. (b) The time evolution of $\Delta S$,
    the excess entropy per site
    of the subsystems, as measured from the entropy per site of the
    whole system. This measures the additional entropy obtained due to
    tracing of degrees of freedom. The graphs
  correspond to a system hopping $g=1.0$, a bath hopping
  $t_B=2.0$, a system bath coupling $\epsilon=0.2$. The bath
  temperature $T=1.0$ and chemical potential $\mu=-1.11$, so that the
  equilibrium density of the system is $\rho_{eq}=0.32$ in this case.}
  \label{Fig:Neqbm_Ent_sizedep}
\end{figure}

As this open
quantum system evolves, we track the dynamics of entanglement entropy
in the system with the formalism we have developed. There are three distinct processes that happen
and leave their imprint on both the correlation functions and
entanglement measures in the system. The first is an exchange of
particles between the system and the bath, which changes the imposed
density pattern in the system. This is a local and incoherent
process depicted in figure \ref{Fig:Neqbm_density}(a). The second process corresponds to creation of two domain
walls at the center which spread outwards like a wavefront, leading to
a domain of particles going right and a domain of holes going
left. This is schematically depicted in \ref{Fig:Neqbm_density}(c). These wavefronts are then reflected back from the boundaries
towards the center. This is equivalent to sloshing of particle and
hole domains in the system and is a coherent quantum process. The waves are damped due to the incoherent
exchange with the bath and eventually die off. The third process that
happens is the creation and annihilation of additional domain walls
due to hopping of the particles, resulting in local Rabi oscillations in the system. This mechanism, depicted in Fig \ref{Fig:Neqbm_density}(d) is also a
coherent quantum process. 

Let us first focus on the evolution of the density profile in this
system. In Fig~\ref{Fig:Neqbm_density}(e), we plot the particle density as a
function of space and time in a color plot. The wavefront
dynamics is clearly visible as a sharp jump in the density profile,
which propagates outward from the center with a uniform velocity. On
the right hand section of the lattice, which starts with no particles,
the jump leads to an increase in the density, while it leads to a
decrease in density on the left half which starts with a local density
of $1$. This is equivalent to a particle and a hole domain moving
outward. The reflection of the waves from the boundary leads to the
characteristic diamond shape in the figure. The creation and annihilation of domain
walls create subsidiary wavefronts, leading to the additional ripples
seen in the pattern. Finally the settling down of the wave is due to
damping coming from interaction with the bath.   We note that the
wavefront velocity $v \sim 1.9~ga$, whereas the subsequent ripples give
a timescale $\tau_r \sim g^{-1}$, as expected from the hopping scale
in the problem. This is clearly seen in Fig.~\ref{Fig:Neqbm_density}(f), where we
plot the time evolution of density of a single site on the left and right half of the
system. The sharp jumps correspond to the wavefront passing through
the site, while the oscillations, which occur after the initial
wavefront has passed, have a frequency $\sim g$. The damping of the coherent
motion occurs on the dissipation time scale provided by the bath, $\tau_{B} \sim
2t_B/\epsilon^2 = 20$, as seen from Fig.~\ref{Fig:Neqbm_density} (e) and (f).

We now shift our focus to the dynamics of entanglement measures in
this system. In Fig.~\ref{Fig:Neqbm_Ent_2site}(a), we plot the von-Neumann
entanglement entropy per site of several 2 site subsystems of our open
quantum system as a function of time. These two sites ($[i,i+1]$) are next to each
other, but their location is varied from the center outward towards
the right. Defining $\delta=i-L/2$, i.e. the distance from the center,
the curves correspond to $\delta=0$ (blue), $\delta=2$ (orange),
$\delta=4$ (green), $\delta=6$ (red), and $\delta=8$ (purple), as
shown in Fig.~\ref{Fig:Neqbm_Ent_2site}(a).
The entropy starts at zero, as expected for a product
state, and there is an initial rise which is independent of the
location of the subsystem. This is dominated by the exchange of
particles with the bath and consequent hopping of these particles
between the two sites. Then, there is a sharp rise in the entanglement
entropy of the subsystem when the wavefront of the domain sloshing
passes through the subsystem. This sharp rise happens at a later time
as we move the subsystem outwards from the center, and the time of the
jump coincides with the time when the average density in the subsystem also
shows a jump, as shown in Fig.~\ref{Fig:Neqbm_Ent_2site}(c) \& (d). We can thus correlate this feature with the passing of
the wavefront. The entanglement entropy then rises slowly with a
shoulder like feature, which has oscillations superimposed on
it. During this time the system is undergoing local Rabi oscillations
between nearest neighbours leading to creation and annihilation of
additional domains in the system. The incoherent exchange with the
bath is also active during this time. One can again see a sharp jump
around $t=20$, when the reflected wave passes through the
subsystem. Note that this jump occurs first for subsystems farther from
the center, since the reflected wave reaches this point
earlier. Beyond this point, the entanglement entropy settles into a
long time decay to its final equilibrium value. During this time, the
entanglement entropy is independent of the location of the subsystem,
since the local incoherent exchange with the bath is dominating the
dynamics during this period.

We have seen that there is a clear correlation between features in the
dynamics of density and entanglement entropy of the system. An obvious
question is whether one can explain the dynamics of
the entanglement entropy solely in terms of the density dynamics. To
explore this question, note that for a spinless system, fixing the
density of a $1$ site subsystem fixes its density matrix, and hence its
entanglement entropy. If the density is $\rho$, this local
entanglement entropy $S_{\text{loc}}=-[\rho \ln \rho + (1-\rho)\ln
(1-\rho)]$. Thus, we can compare the entanglement entropy of the two
site subsystem with that of an effective one site subsystem with the
same average density. The difference will be related to entanglement
between the two sites making up the subsystem.
In Fig~\ref{Fig:Neqbm_Ent_2site}(c), we plot the entanglement
entropy and average density of a $2$-site subsystem starting at the
$24^{th}$ site (this is to the
right of the center) as a function of time. In the same figure, we also
plot the entropy of an effective $1$-site subystem with average occupation
$\rho(t)$, i.e.
\beq
S_{\text{loc}}(t) =-[ \rho(t) \ln \rho(t) + (1-\rho(t)) \ln
(1-\rho(t))].
\eeq
We see that these two curves fall on top of each other
till the first coherent wave hits the subsystem, and then they
differ from each other, while following the same general
trends. Similar trends can be seen for another two site subsystem starting on the
$35^{th}$ site in Fig~\ref{Fig:Neqbm_Ent_2site}(d). In this case, the
coherent wave passes later and hence the entanglement entropy follows
$S_{\text{loc}}$ for a longer time. This reconfirms the idea that the initial
dynamics is incoherent till the domain wall wavefront hits the subsystem.

A better way to distinguish the entanglement between the degrees of
freedom in the subsystem is to calculate the mutual information. For the
two site subsystem composed of site $i$ and site $i+1$, the mutual information between the sites, $I$ is given by
\beq
I(i,i+1)= S_{\text{vN}}(i,i+1) -S_{\text{vN}}(i) -S_{\text{vN}}(i+1),
\eeq
where $S_{\text{vN}}(i,i+1)$ is the entanglement entropy of the two state
system, $S_{\text{vN}}(i)$ and $S_{\text{vN}}(i+1)$ are the entanglement of the
corresponding single site systems. This quantity would be zero if the
two sites are not entangled. The mutual information between the two
consecutive sites at different locations are plotted as a function of time in
Fig~\ref{Fig:Neqbm_Ent_2site}(b). We have plotted the mutual information for the same
set of subsystems for which entanglement entropy was plotted in
Fig~\ref{Fig:Neqbm_Ent_2site}(a) with the same color coding.  In this case, we find large jumps with oscillations
superimposed on an increasing background. The background is
independent of the location of the subsystem. This comes from the
change in particle density due to exchange with the bath, and these
particles getting entangled by hopping. The initial jumps coincide with the
passing of the coherent wavefront and occurs later for subsystems
which are located farther from the center. This is shown in the inset
of Fig~\ref{Fig:Neqbm_Ent_2site}(b) for small times. The mutual information decays with some
oscillations (created by breaking up of the domain walls) and finally
settles to the background once this wave is damped around $t=20$. The
mutual information thus cleanly picks up the quantum coherence on top of the
increasing background due to incoherent exchange with the bath and
subsequent coupling of the sites due to hopping.

We now focus on how the evolution of the entanglement entropy is
affected by the size of the subsystem. For this we consider subsystems
of increasing size centered around the middle of the system. The time
evolution of the 
entanglement entropy density (i.e. the
entanglement entropy divided by the subsystem size) of different sized subsystems is plotted in
Fig~\ref{Fig:Neqbm_Ent_sizedep}(a). The different sizes plotted are
$L_A=1$ (blue), $L_A=2$ (orange), $L_A=4$ (green), $L_A=8$ (red),
$L_A=16$ (purple), and $L_A=40$ (brown), which corresponds to the
whole system. While the entanglement entropy of the subsystems show
the sudden jump and oscillations evident in the two site systems, the
amplitude of these oscillations go down as we look at larger and
larger subsystems. In fact the evolution of entropy for the full system is smooth
and devoid of these features. The finite subsystems follow this curve
upto a point and then deviate to manifest the effects of quantum processes
in the system.

When we calculate the entanglement entropy of a subsystem in a pure quantum state, it
has a simple interpretation: the loss of information about the quantum
degrees of freedom shows up as entropy of the reduced density
matrix. However in an open quantum system, the whole system evolves
from a pure state to a density matrix and has an entropy of its own
(the $L=40$ curve shows evolution of this entropy). To take this into account, we
define the excess entropy density of a subsystem,
\beq
\Delta S= \frac{S_{\text{vN}}(L_A)}{L_A}-\frac{S_{\text{vN}}(L)}{L}.
\eeq
This is the additional randomness introduced into the subsystem due to
tracing of the complementary degrees of freedom. In
Fig~\ref{Fig:Neqbm_Ent_sizedep}(b), we plot the time evolution of
$\Delta S$ for different subsystem sizes (with same color coding as
Fig~\ref{Fig:Neqbm_Ent_sizedep}(a)). We see that this excess entropy
shows a steep jump followed by a decay with oscillations superposed on
it. The characteristic jumps due to passing of the wavefront is
clearly evident in this plot. We also note that in the long time
limit, it is clear that the smaller subsystems have more excess
entropy density. This is expected since we are tracing over larger
number of degrees of freedom in this case, leading to more information loss.

\section{Conclusions\label{Conclusions}}

In this paper, we have formulated a new way of calculating entanglement
entropy of Fermionic systems through the construction of a Wigner
functio,n  which is a Grassmann valued function of
Grassmann variables. The Wigner function is then identified
with the Keldysh partition function of the system with a set of
sources, which are proportional to the arguments of the Wigner
function.

We have extended this formalism to non-equilibrium dynamics starting
from arbitrary initial conditions. For a non-interacting fermionic
open quantum system, starting from an initial Fock state, we have
derived exact formulae for the entanglement entropy of a
subsystem. This is the key new universal result in this paper, which
has a wide scope of application in different situations.

We have used our formalism to look at entanglement entropy of momentum
Fock states of one-dimensional Fermions. We find that the states can be classified by
the number of contiguous blocks of momentum occupancy in them. This is
also related to the number of zeroes in the dispersion of the
effective Hamiltonian, for which this Fock state is a ground state. If
the number of momentum occupancy blocks is smaller than the size of
the subsystem, the entanglement entropy of the subsystem scales
logarithmically with the subsystem size, and the state looks
``critical''; i.e. shares the property of many body systems at phase transitions. On the other hand when the number of blocks is larger
than the subsystem size, the entanglement entropy scales linearly with
the subsystem size, which is a typical property of thermal
systems. So, the same state can either look ``critical'' or
``thermal'' depending on the range of subsystem size one is looking
at. We use this idea to analytically estimate the number of
``critical'' states for a given subsystem size.

Finally, we use our formalism to study the evolution of entanglement
entropy of subsystems of a one dimensional open quantum system, which is initialized
to a state with a domain wall at the center of the lattice. We
understand the dynamics in terms of the coherent motion of the domain
walls together with incoherent exchange of particles with the bath.

We would like to note that the formulae that we have derived in this
paper are applicable to a large class of systems under different
situations. They will especially help in understanding behaviour of
entanglement entropy in higher dimensional systems, where there are
very few answers known, but we leave this question for a future work,

\appendix
\section{Wigner Characteristic and Diagrammatic Expansion of R\'{e}nyi
	Entropy\label{diag}}

\begin{figure}[t!]
	\includegraphics[width =\columnwidth]{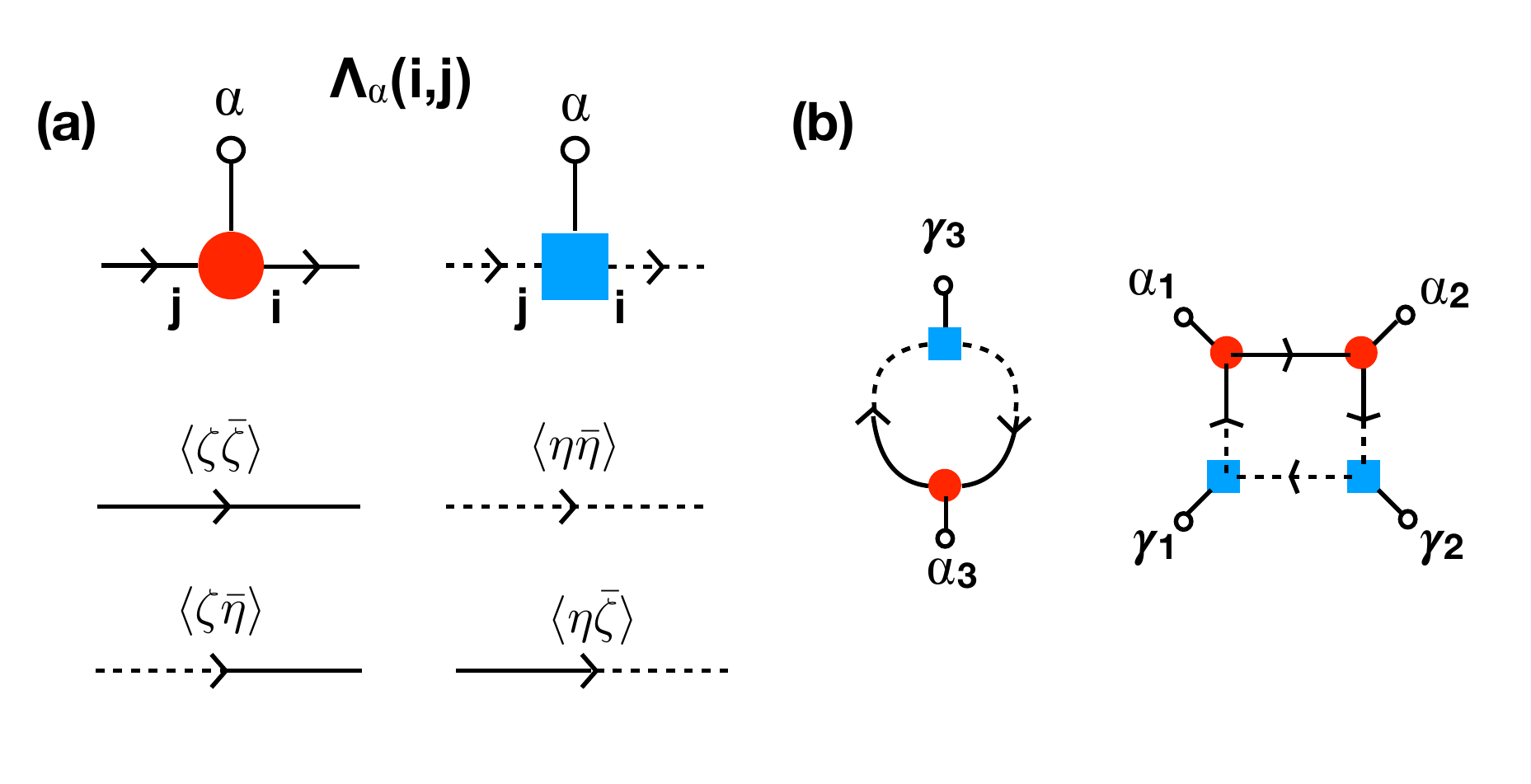}
	\caption{(a) Vertices and propagators used in evaluating the second
		R\'{e}nyi entropy. Note that the solid lines begin and end at 
		circular vertices, while the dotted lines begin and end at the
		square vertices. The third line from the vertices (ending in a
		small circle) is used to denote the mode or $\alpha$ index of the
		vertex. The lines coming from the vertices are joined to form
		propagators. The four kinds of propagators are indicated in the
		figure. (b) A disconnected diagram for the evaluation of
		$\mathrm{e}^{-S^{(2)}}$. Note that the indices of the circular vertices are
		constrained to be $\alpha_1\neq\alpha_2\neq\alpha_3$. Similarly,
		the indices of the square vertices must satisfy
		$\gamma_1\neq\gamma_2\neq\gamma_3$. These constrained summations
		make evaluation of these diagrams difficult.}
	\label{Fig_vert_prop}
\end{figure}
In this section, we will provide an alternative derivation of the
formulae derived above, which will also indicate a way forward for the
case where the number of particles is smaller than the subsystem
size. In this case, we will first obtain the Wigner characteristic
function by taking the derivatives with respect to initial sources in
Eq.~\ref{Wigchar_rho} to write
\begin{equation}
\chi_D(\ve{\zeta},\ve{\zbar},t) =\mathrm{e}^{-\frac{1}{2} \ve{\zbar}
	\hat{\Gamma}(t)\ve{\zeta}}\prod_\alpha [1+n_\alpha \ve{\zbar} \lhat^\alpha(t) \ve{\zeta}]
\end{equation}
where $\ve{\zbar}=(\zbar_{x_1},\zbar_{x_2} ... \zbar_{x_{V_A}})$, and
$\hat{\Gamma}$ and $\lhat$ are matrices in $x$,$x'$ space. The second
R\'{e}nyi entropy is then given by
\begin{widetext}
	\begin{equation}
	\mathrm{e}^{-S^{(2)}}=\int {\cal D}[\ve{\zeta},\ve{\zbar}]{\cal D}[\eta,\ebar] \mathrm{e}^{-\frac{1}{2} \left(\bar{\ve{\zeta}},\bar{\ve{\eta}}\right)\left(\begin{array}{cc}
			\hat{\Gamma}(t) &  -\hat{1}\\
			\hat{1} & \hat{\Gamma}(t) \end{array}\right) \left(\begin{array}{c}
			\ve{\zeta}\\
			\ve{\eta}\end{array}\right)}\prod_\alpha
	[1+n_\alpha \ve{\zbar} \lhat^\alpha(t)
	\ve{\zeta}]\prod_\gamma [1+n_\gamma \ve{\ebar}
	\lhat^\gamma(t) \ve{\eta}]
	\label{S2_Wick}
	\end{equation}
      \end{widetext}
where $n_\alpha$ is the occupation number of the mode $\alpha$ in the
initial state. The appearance of integrals of a gaussian function multiplied by
polynomials in Eq.~\ref{S2_Wick} suggest the use of Wick's theorem and resultant
diagrammatic representations to evaluate $\mathrm{e}^{-S^{(2)}}$. The key
elements of this expansion are shown in Fig. ~\ref{Fig_vert_prop}. 
%\begin{figure}[h!]
%	\includegraphics[width =\columnwidth]{Vertices_Prop.pdf}
%	\caption{(a) Vertices and propagators used in evaluating the second
%		R\'{e}nyi entropy. Note that the solid lines begin and end at 
%		circular vertices, while the dotted lines begin and end at the
%		square vertices. The third line from the vertices (ending in a
%		small circle) is used to denote the mode or $\alpha$ index of the
%		vertex. The lines coming from the vertices are joined to form
%		propagators. The four kinds of propagators are indicated in the
%		figure. (b) A disconnected diagram for the evaluation of
%		$\mathrm{e}^{-S^{(2)}}$. Note that the indices of the circular vertices are
%		constrained to be $\alpha_1\neq\alpha_2\neq\alpha_3$. Similarly,
%		the indices of the square vertices must satisfy
%		$\gamma_1\neq\gamma_2\neq\gamma_3$. These constrained summations
%		make evaluation of these diagrams difficult.}
%	\label{Fig_vert_prop}
%\end{figure}
The
vertices $n_\alpha \lhat^\alpha$ coupling to the $\ve{\zbar},\ve{\zeta}$
variables is represented by a circle with three legs sticking out;
the top leg ending in a small circle indicates the label $\alpha$,
which we will refer to as ``index'' of the vertex. The horizontal outgoing leg
indicates the coupling to $\ve{\zbar}$ and  the horizontal incoming leg
indicates the coupling to$\ve{\zeta}$. Similar construction is done with square
vertices for $n_\gamma \lhat^\gamma$ s coupling to $\ve{\ebar},\ve{\eta}$
variables. Note that the $\ve{\zbar},\ve{\zeta}$ variables are denoted by thick
lines, while $\ve{\ebar},\ve{\eta}$ variables are denoted by dashed lines. While
the square and circular vertices represent the same matrix, it is
useful to keep them as separate vertices for various book keeping
purposes. These are shown in
Fig. ~\ref{Fig_vert_prop}(a). Fig.~\ref{Fig_vert_prop} (a) also shows the propagators: the expectation $\langle \zeta\zbar\rangle =2\left
[\hat{1}+\hat{\Gamma}^2\right]^{-1}\hat{\Gamma}$ is denoted by a thick
straight line, $\langle \eta\ebar\rangle =2\left
[\hat{1}+\hat{\Gamma}^2\right]^{-1}\hat{\Gamma}$ by a dashed straight
line, $\langle \eta\zbar\rangle =-2\left
[\hat{1}+\hat{\Gamma}^2\right]^{-1}$ is denoted by a dashed-thick line
and $\langle \ebar\zeta\rangle =2\left
[\hat{1}+\hat{\Gamma}^2\right]^{-1}$ is denoted by a thick-dashed
line. The diagrams for $\mathrm{e}^{-S^{(2)}}$ then correspond to motifs where no lines are
hanging out (much like diagrams for partition functions in standard
field theories). They are composed of rings with square and circular
vertices sitting on them. Note that a diagram for $\mathrm{e}^{-S^{(2)}}$ can
consist of several such disconnected rings. One such diagram is shown
in Fig. Fig. ~\ref{Fig_vert_prop} (b).
      \begin{figure*}[t!]
	\includegraphics[width=\columnwidth]{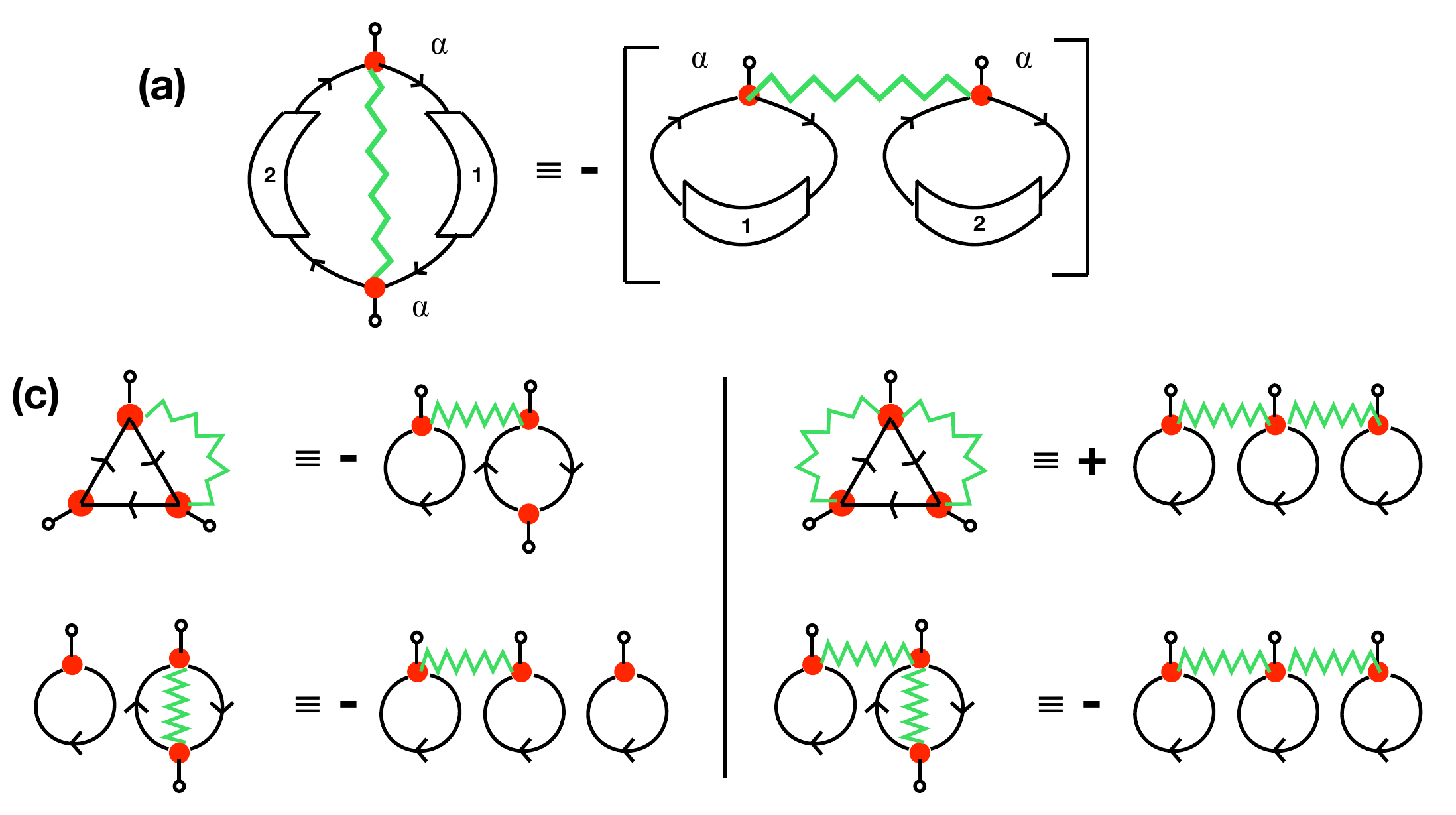}
	\includegraphics[width =\columnwidth]{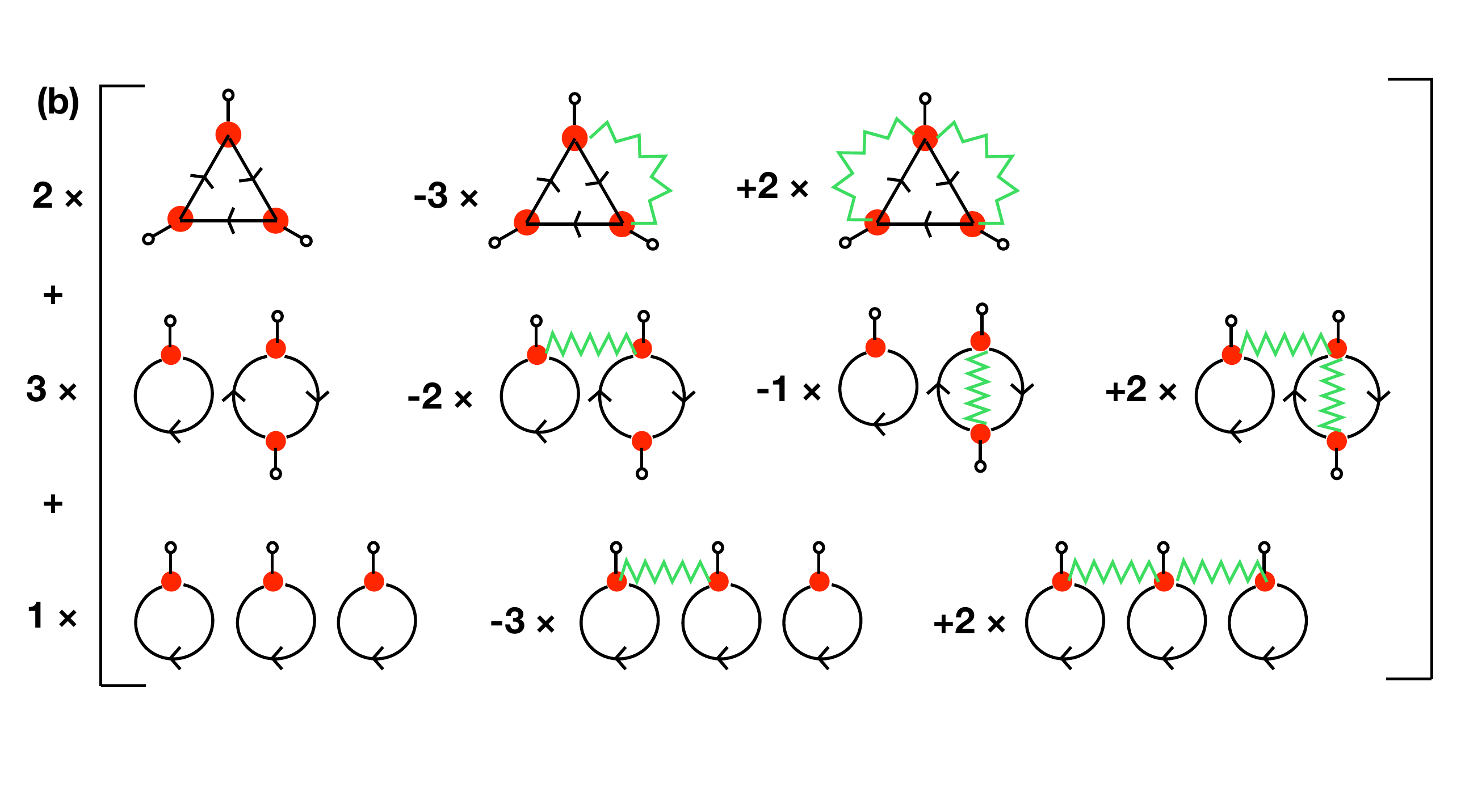}
	\caption{(a) Vertices and propagators used in evaluating the second
		R\'{e}nyi entropy. Note that the solid lines begin and end at 
		circular vertices, while the dotted lines begin and end at the
		square vertices. The third line from the vertices (ending in a
		small circle) is used to denote the mode or $\alpha$ index of the
		vertex. The lines coming from the vertices are joined to form
		propagators. The four kinds of propagators are indicated in the
		figure. (b) A disconnected diagram for the evaluation of
		$\mathrm{e}^{-S^{(2)}}$. Note that the indices of the circular vertices are
		constrained to be $\alpha_1\neq\alpha_2\neq\alpha_3$. Similarly,
		the indices of the square vertices must satisfy
		$\gamma_1\neq\gamma_2\neq\gamma_3$. These constrained summations
		make evaluation of these diagrams difficult.}
	\label{Fig_cancel}
\end{figure*}
The rules for evaluating a particular diagram can be given by: (i) For each vertex put the
corresponding $n_\alpha\hat{\Lambda}_\alpha$ matrix (in $x$,$x'$ space). (ii) For
each propagator multiply by the corresponding matrix; keep the order
of multiplication intact since $\hat{\Gamma}$ and $\hat{\Lambda}_\alpha$ do not
commute when there is a Keldysh self-energy due to coupling to an
external bath. (iii) For each ring, take the trace of this product in the $x$, $x'$
space and multiply.  (v) Multiply by $(-1)^F$, where $F$ is the number of
Fermion loops (in this case, number of disconnected rings). (iv) For each diagram, multiply by the symmetry factor,
which is the number of different connections which produce the same
diagram.
 (vi) Sum
over all possible $\alpha$ s and $\gamma$ s of the $\Lambda$ matrices,
making sure that all $\alpha$ values are distinct and all $\gamma$
values are distinct. (vii) This last constraint comes from expanding the product 
\begin{widetext}
	\begin{equation}
	\prod_{\alpha} \left[1+ n_\alpha \zbar_x \hat{\Lambda}^\alpha(xx')\zeta_{x'}\right]= 1+\sum_\alpha n_\alpha \zbar_x \hat{\Lambda}^\alpha(xx')\zeta_{x'}
	+\frac{1}{2!} \sum_{\alpha\neq\beta} n_\alpha \zbar_{x_1}
	\hat{\Lambda}^\alpha(x_1x'_1)\zeta_{x'_1} n_\beta \zbar_{x_2}
	\hat{\Lambda}^\beta(x_2x'_2)\zeta_{x'_2} +\cdots .
	\end{equation}
\end{widetext}
The constrained sum leads to great
difficulties in evaluating the diagrams; since one cannot treat the
$\alpha$ and $\gamma$ indices as internal indices to be summed over
independently, one has to evaluate the diagrams resulting from all the
permutations of these indices separately, leading to exponential
growth in number of diagrams. 

We note that a-priori there is no small
parameter in this expansion and hence it does not make sense to
evaluate a few diagrams.
\begin{figure}[h!]
	\includegraphics[width=0.95\columnwidth]{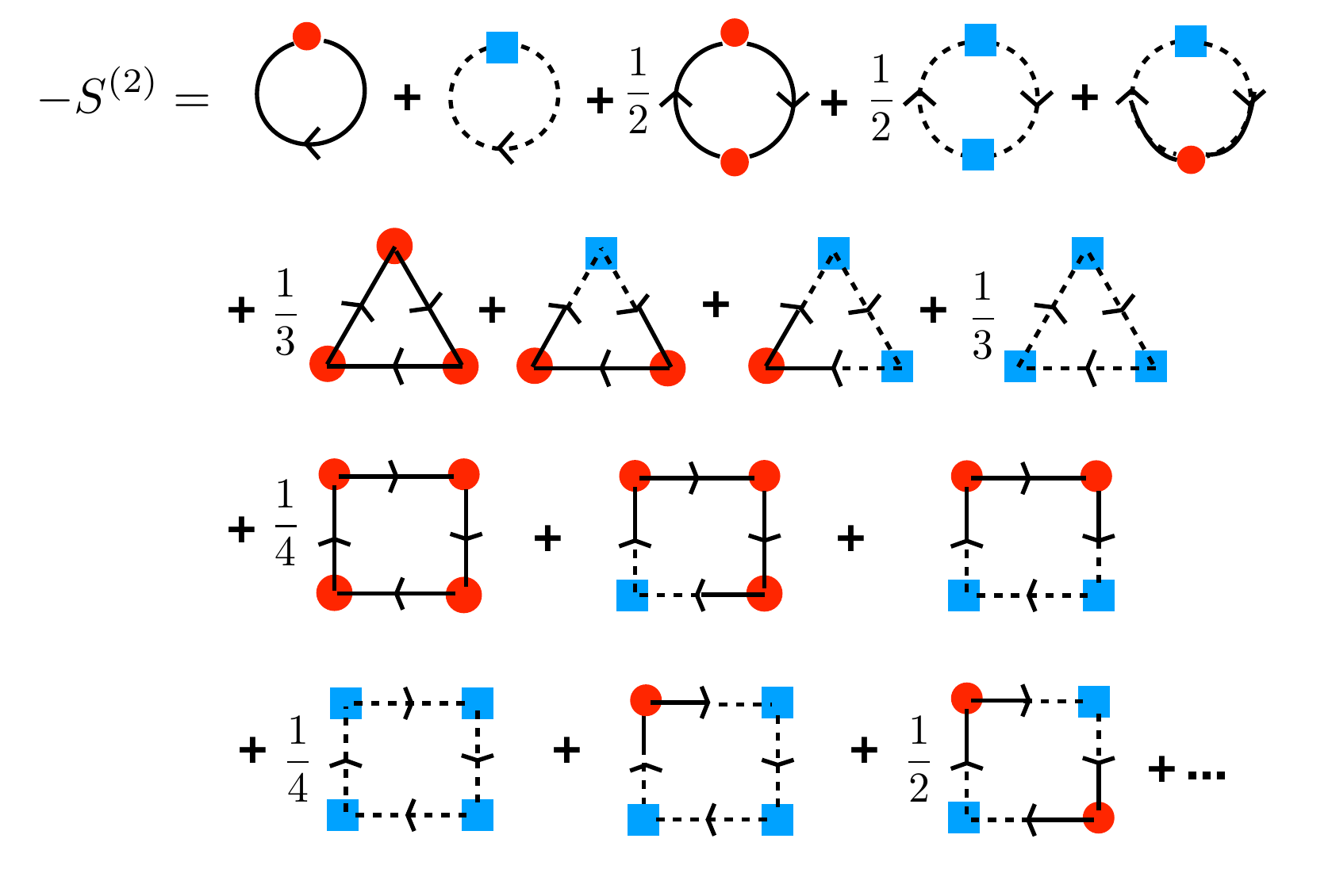}
	\caption{The series of connected diagrams for calculation of
		$S^{(2)}$}
	\label{Fig_Connected}
\end{figure}
However, for the case of Fermions, one can
get rid of the constrained summations by the following argument:
The constrained sums over distinct pairs can be written in terms of unconstrained sums
and identification of variables, e.g. $\sum_{(a,b)}f(a,b)=\frac{1}{2!}
\left[\sum_{ab} f(a,b)-\sum_a f(a,a)\right]$ and so on, where $(a,b)$ denote
distinct pairs. Thus the constrained summation can be overcome at the
expense of having additional diagrams, where some of the circular (or
square) vertices are identified. We denote such an identification by
putting a wavy line joining the vertices. Fig. ~\ref{Fig_cancel} (a)
shows all diagrams (with associated symmetry factors) with three
circular vertices in the expansion of $\mathrm{e}^{-S^{(2)}}$, written in terms
of identified vertices. Note that diagrams can have more than two
vertices identified.

The matrix
$n_\alpha\hat{\Lambda}_\alpha$ has a factorizable form, i.e. $\hat{\Lambda}
^\alpha(x,x')
\sim g_\alpha(x)g^\ast_\alpha(x')$ and $n_\alpha^2=n_\alpha$. Using this, one can then easily show that $\tr{
\hat{\Lambda}^\alpha \hat{A} \hat{\Lambda}^\alpha\hat{B}}= \left(\tr{\hat{A}
\hat{\Lambda}^\alpha}\right) \left(\tr{\hat{B}\hat{\Lambda}^\alpha}\right)$ for any
matrices $\hat{A}$ and $\hat{B}$. In terms of diagrams, this implies
that a ring, where two indices are identified, is equal to negative of a
``disconnected'' diagram with two rings each involving one of the
indentified vertices. This is shown in  Fig. ~\ref{Fig_cancel}
(b). The negative sign comes because the equivalent ``disconnected''
diagram has an extra fermion loop, while the Trace factorization does
not give any minus sign. With this, one can show that diagrams
with any non-zero number of identifications cancel each other. For example, the
equivalences between the diagrams in Fig. ~\ref{Fig_cancel} (a)  are
shown in Fig. ~\ref{Fig_cancel} (c). It is easy to see from a simple
counting that the diagrams in Fig. ~\ref{Fig_cancel} (a) with the wavy
lines sum to zero. This cancellation works out for each set of 
diagrams which has $n$ circular vertices and $m$ square vertices
(these include disconnected diagrams of a particular ``order''). Note
that the Fermion minus sign is crucial for this cancellation, and this
would not occur for a similar construction with Bosons.

The net result of the cancellation described above is that we can
simply replace the ``distinct pair summations'' over the indices by
unconstrained summations. Since the propagators do not depend on the
index, this allows us to replace every vertex by $\hat{\Lambda}
=\sum_\alpha n_\alpha \hat{\Lambda}^\alpha$, and hence we can now drop
the line on the vertices ending in a small circle, which was used to
indicate the index. Going back to the integral which gave rise to this
diagrammatics, we can now write it as
\begin{widetext}
	\begin{equation}
	\mathrm{e}^{-S^{(2)}}=\int {\cal D}[\ve{\zeta},\ve{\zbar}]{\cal D}[\ve{\eta},\ve{\ebar}]~ \mathrm{e}^{-\dfrac{1}{2}
	\begin{matrix}
	\begin{pmatrix}
		\ve{\bar{\zeta}}&\ve{\bar{\eta}}
	\end{pmatrix}\\
	\hspace{2pt}
	\end{matrix} %
\left(\begin{array}{cc}
			\hat{\Gamma}(t) &  -\hat{1}\\
			\hat{1} & \hat{\Gamma}(t) \end{array}\right) \left(\begin{array}{c}
			\ve{\zeta}\\
			\ve{\eta}\end{array}\right)}\sum_{l=0}^N
	\frac{\left[ \ve{\zbar} \lhat(t)
		\ve{\zeta}\right]^l}{l!}\sum_{m=0}^N
	\frac{\left[ \ve{\ebar} \lhat(t)
		\ve{\eta}\right]^m}{m!}
	\label{S2_lambda}
	\end{equation}
	We note that the number of Grassmann variables $\zeta_x$ (and
	$\eta_x$) is $L_A$. Hence for $l>L_A$, $\left[ \ve{\zbar}
	\lhat(t)\ve{\zeta}\right]^l$ inevitably repeats one or more
	$\zeta_x$ s in the string, and the string is then zero by
	definition. For $N \geq L_A$, this allows us to extend the summation
	over $l$ or $m$ in Eq.~\ref{S2_lambda} to $\infty$ and we get a factor
	of $\mathrm{e}^{\ve{\zbar}
		\lhat(t)\ve{\zeta}+\ve{\ebar}
		\lhat(t)\ve{\eta}}$ from the polynomials. Combining with the
	Gaussian part, this gives
	\begin{equation}
	\mathrm{e}^{-S^{(2)}}=\int {\cal D}[\ve{\zeta},\ve{\zbar}]{\cal D}[\ve{\eta},\ve{\ebar}]~ \mathrm{e}^{-\dfrac{1}{2}
		\begin{matrix}
			\begin{pmatrix}
				\ve{\bar{\zeta}}&\ve{\bar{\eta}}
			\end{pmatrix}\\
			\hspace{2pt}
		\end{matrix} %
	\left(\begin{array}{cc}
			\hat{\Gamma}(t)-2\hat{\Lambda}(t) &  -\hat{1}\\
			\hat{1} & \hat{\Gamma}(t) -2\hat{\Lambda}(t)\end{array}\right) \left(\begin{array}{c}
			\ve{\zeta}\\
			\ve{\eta}\end{array}\right)}
	\label{S2_lambda_gauss}
	\end{equation}
	which leads to the formula Eq.~\ref{S2_final}.                                 
\end{widetext}

Further, if we assume that the number of particles
$N$ goes to infinity, the series does not terminate after a finite
number of terms. Then we can use the linked cluster theorem to show
that the expansion for $S^{(2)}$ will only have connected diagrams,
i.e. single rings with different arrangement of square and circular
vertices. 
%\begin{figure}[b!]
%	\includegraphics[width=0.95\columnwidth]{Connected_Diag.pdf}
%	\caption{The series of connected diagrams for calculation of
%		$S^{(2)}$}
%	\label{Fig_Connected}
%\end{figure}
The first few terms in this series (upto $4^{th}$ order in $\hat{\Lambda}$ has been shown in
Fig.~\ref{Fig_Connected}. The diagram evaluation rules are same as
before with two additional rules:  (a) vertices are represented by
$\hat{\Lambda}$ and (b) an
additional factor of $1/{m!n!}$ multiplies diagrams with $m$ circular and $n$
square vertices. 

% Let us now consider the case of a closed system. In this case, there
% is no Keldysh self energy and
% \beq
% \Gamma(x,x',t) = \sum_\alpha
% G^R(x,t;\alpha,0)[G^R(x',t;\alpha,0)]^\ast =\delta_{x,x'}
% \eeq
% This simplifies the evaluation of the diagrams, since $\langle
% \zeta\zbar\rangle =\langle \eta\ebar\rangle =\langle \zeta\ebar\rangle
% =1$ and $\langle \eta\zbar\rangle=-1$. If one considers a diagram
% with $n$ circular vertices, $m$ square vertices, and $p$ links between
% square and circular vertices (i.e. $p$ $\langle \eta\zbar rangle$ ), this diagram will contribute
% $(-1)^{p+1}\frac{1}{m!n!}~Tr ~[-2\hat{\Lambda}]^{m+n}$ where the extra
% $-$ sign is due to the Note that the number $p$ of $\langle
% \zbar \eta\rangle$ and $\langle \ebar \zeta$ links appearing in a ring
% has to be the same and $p$ is bounded from above by $min(m,n)$.
% The
% symmetry factor for such a diagram is given by $(m-1)! (n-1)!
% \left.^{m}C_{p}\right. \left.^{n}C_{p}\right. p$ 
% for $p>0$, while for $p=0$, i.e. diagrams containing only $m$ square or $m$
% circular vertices, this is given by $(m-1)!$. Finally, working
% out the summation over $p$, we get the series
% \beq
% S^{(2)}= \sum_{m>0} \frac{2}{m} Tr ~ \hat{\Lambda}^m - \sum_{m,n
%   >0}\left._2F_1[1-m,1-n,2,-1]Tr ~ \hat{\Lambda}^{m+n} \right.
% \eeq
% We have checked upto very high orders that this series is equivalent
% to 
% \beq
% S^{(2)}= - Tr ~Log\left[\hat{1}+( \hat{1}-2\hat{\Lambda})^2\right]
% \eeq

\section{Derivation of Formulae for $S^{(n)}$ and
	$S_{vN}$}\label{app:Sn}

In this Appendix, we show the derivation of the formulae for $S^{(n)}$
which has been quoted in the main text. The proof follows the essence of the analysis in section \ref{ExactFormula}; we start from the expression of $S^{(n)}$ in terms of the characteristic function given in Eq.\ref{eqref:Sngen}, evaluate it for a generic free field theory to get an expression for $\tr{\hat{\rho}^n}$ in presence of the additional sources $u$. Then we act with the derivative operators $\mathcal{L}$ to get a determinant involving the physical Keldysh Greens function at equal times, $\mathcal{G}^{K}$, which is simplified to get Eq.\ref{eqnref:Sn}. 

Substituting $\chi^r_D$ from Eq.\ref{chidu} into the expression for $\tr{\hat{\rho}_r^{n+1}}$ as inferred from Eq.\ref{eqref:Sngen}, we get
\begin{widetext}
	\begin{equation}
	\begin{split}
	\tr{\hat{\rho}_r^{n+1}}(\ve{u},\{\ve{v}^{(i)}\})=2^{nV_A}%
	\int  \prod_{i=1}^{n}%
	\mathcal{D}[\ve{\zeta}^{(i)},\ve{\zbar}^{(i)}]%
	\mathcal{D}[\ve{\eta}^{(i)},\ve{\bar{\eta}}^{(i)}]%
	\exp\frac{1}{2}\left[-\sum_i\ve{\bar{\eta}}^{(i)}\ii\hat{G}^{K}(\ve{v}_i)\ve{\eta}^{(i)}%
	-\sum_{ij}\ve{\zbar}^{(i)}\ii\hat{G}^{K}(\ve{u})\ve{\zeta}^{(j)}%
	\right]\\%
	\exp\frac{1}{2}\left({\sum_{i}\ve{\zbar}^{(i)}\!\cdot\ve{\eta}^{(i)}-\ve{\bar{\eta}}^{(i)}\!\cdot\ve{\zeta}^{(i)}%
		+\sum_{i>j}\ve{\zbar}^{(i)}\!\cdot\ve{\zeta}^{(j)}-\ve{\bar{\zeta}}^{(j)}\!\cdot\ve{\zeta}^{(i)}}\right)
	\end{split}
	\label{eqref:rhon}
	\end{equation}
	
	where $\ve{u}$ and $\ve{v}^{(i)}$ are the auxiliary sources keeping track of initial condition $\hat{\rho}_0$. The corresponding R\'{e}nyi entropy is recovered after acting on the above with a string of differential operators $\mathcal{L}$, one for each instance of $\chi_D^r$, i.e.,
	\begin{equation}
	\mathrm{e}^{-(n)S^{(n+1)}}:=\mathcal{L}(\partial_{\ve{u}},\hat{\rho}_0)\prod_{i=1}^{n}\mathcal{L}\left(\partial_{\ve{v^{(i)}}},\hat{\rho}_0\right)\tr{\hat{\rho}_r^{n+1}}(\ve{u},\{\ve{v}^{(i)}\})\label{eqref:L}
	\end{equation}
	It is convenient to re-imagine the fields $\ve{\zeta}^{(i)}$ to be components of $\ve{Z}=(\ve{\zeta}^{(1)},\ve{\zeta}^{(2)},\cdots,\ve{\zeta}^{(n)})^{T}$ and $\eta^{(i)}$ as components of $\ve{E}=(\ve{\eta}^{(1)},\ve{\eta}^{(2)},\cdots,\ve{\eta}^{(n)})^{T}$, and similarly for the bar-ed fields. Re-expressing Eq.\ref{eqref:rhon} in terms of objects in this superspace of $n$ copies of the fields, 
	%  \begin{equation}
	\begin{align}
	\tr{\hat{\rho}_r^{n+1}}(\ve{u},\{\ve{v}^{(i)}\})&=%
	2^{nV_A}%
	\int %
	\mathcal{D}[\ve{Z},\bar{\ve{Z}}]%
	\mathcal{D}[\ve{E},\bar{\ve{E}}]%
	\exp\left[-\frac{1}{2}%
	\begin{matrix}
	\begin{pmatrix}
	\bar{\ve{Z}},\bar{\ve{E}}
	\end{pmatrix}\\
	\hphantom{-}
	\end{matrix}
	\begin{pmatrix}
	\mathbb{I}_F\otimes\ii \hat{G}^{K}(\ve{u})-\mathbb{I}_S\otimes\hat{1} & -\mathbb{I}_D\otimes\hat{1}\\
	\mathbb{I}_D\otimes\hat{1} & \hat{\mathbb{G}}_D(\{\ve{v}^{(i)}\})
	\end{pmatrix}
	\begin{pmatrix}
	\ve{Z}\\
	\ve{E}
	\end{pmatrix}\right]\no\\
	&=\frac{1}{2^{nV_A}}\det%
	\begin{bmatrix}
	\mathbb{I}_F\otimes\ii \hat{G}^{K}(\ve{u})-\mathbb{I}_A\otimes\hat{1} & -\mathbb{I}_D\otimes\hat{1}\\
	\mathbb{I}_D\otimes\hat{1} & \ii\hat{\mathbb{G}}_D(\{\ve{v}^{(i)}\})
	\end{bmatrix}\label{eqref:detSn}
	\end{align}
	%\end{equation}
\end{widetext}
where $\mathbb{I}_D$ is the $n\times n$ identity matrix, $\mathbb{I}_F$ is a $n\times n$ full matrix with each entry being $1$, and $\mathbb{I}_A$ is an antisymmetric matrix with all the entries above the principal diagonal being $-1$ as shown below
\begin{equation}
\begin{gathered}
\mathbb{I}_D=\begin{pmatrix}
1&&0\\
&\ddots&\\
0&&1
\end{pmatrix}_{n\times n}\quad
\mathbb{I}_F=%
\begin{pmatrix}
1&\cdots&1\\
\vdots&\ddots&\vdots\\
1&\cdots&1
\end{pmatrix}_{n\times n}\\[5pt]
\mathbb{I}_A=%
\begin{pmatrix}
0&-1&\cdots&-1\\
1&\ddots&\ddots&\hphantom{-}\vdots\\
\vdots&\ddots&\ddots&-1\\
1&\cdots&1&0
\end{pmatrix}_{n\times n}
\end{gathered}
\end{equation}
$\ii\hat{\mathbb{G}}_D$ is a ${nV_A\times nV_A}$ block diagonal matrix such that
\begin{equation}
\ii\hat{\mathbb{G}}_D(\{\ve{v}^{(i)}\})=\mathrm{Diag}\left(\ii\hat{G}^K(\ve{v}^{(1)}),\cdots,\ii\hat{G}^K(\ve{v}^{(n)})\right)
\end{equation}
%The matrix within the determinant in Eq.\ref{eqref:detSn} reduces to  that in Eq.\ref{eqref:S2mat} for $n=1$ but h

The expression in Eq.\ref{eqref:S2mat} is a special case of the matrix within the determinant in Eq.\ref{eqref:detSn} for $n=1$, more so since the former doesn't correctly anticipate the structure of the latter. However, they both have a similar layout, in sense that each column depends on only a single family of auxiliary sources $\{\ve{u},\{\ve{v}^{(i)}\}\}$. Owing to this fact, each of the differential operators in Eq.\ref{eqref:L} act on columns independently to give a sum of an exponential number of terms, which can be reconstituted into one term like in Eq.\ref{eqref:reconstitute}. This complicated procedure is symbolically equivalent to replacing the auxiliary source dependent Greens functions %with their physical counterparts i.e. 
as $\ii \hat{G}^K(\ve{u})\to\hat{\Gamma}-2\hat{\Lambda}$ in \ref{eqref:detSn}. Thus, defining $\ii\hat{\mathbb{G}}^K_D=\mathbb{I}_D\otimes(\hat{\Gamma}-2\hat{\Lambda})$, we have,
\begin{align}
\mathrm{e}^{-nS^{(n+1)}}&=\frac{1}{2^{nV_A}}\det%
\begin{bmatrix}
\mathbb{I}_F\otimes(\hat{\Gamma}-2\hat{\Lambda})-\mathbb{I}_A\otimes\hat{1} & -\mathbb{I}_D\otimes\hat{1}\\
\mathbb{I}_D\otimes\hat{1} & \ii\hat{\mathbb{G}}^K_D
\end{bmatrix}\no\\
&=
\frac{1}{2^{nV_A}}\det\left[\mathbb{I}_D\otimes\hat{1}+\mathbb{I}_F\otimes(\hat{\Gamma}-2\hat{\Lambda})^2\right.\no\\
&\hphantom{=
	\frac{1}{2^{nV_A}}\det\mathbb{I}_D\otimes\hat{1}+\mathbb{I}_F}\left.-\mathbb{I}_A\otimes(\hat{\Gamma}-2\hat{\Lambda})\right]\label{detstruct}
\end{align}
where in the last line we have used a result about the determinants of block matrices\cite{silvester2000determinants}. Note that for $n=1$, $\mathbb{I}_S=0,~\mathbb{I}_D=\mathbb{I}_F=1$ and we exactly recover our result for $S^{(2)}$. We now evaluate the determinant in Eq.\ref{detstruct} to get an expression free of the $\mathbb{I}$ structures. First we can show that for matrices $\hat{A},\hat{B}$ such that $[\hat{A},\hat{B}]=0$
\begin{equation}
\begin{gathered}
\det\left[\mathbb{I}_D\otimes\hat{1}+\mathbb{I}_F\otimes(\hat{A}\hat{B})-\mathbb{I}_A\otimes\hat{B}\right]\qquad\qquad\\%
\qquad=%
\det\left[\frac{\hat{1}-\hat{A}}{2}\left(\hat{1}-\hat{B}\right)^n%
+\frac{\hat{1}+\hat{A}}{2}\left(\hat{1}+\hat{B}\right)^n\right]
\end{gathered}
\end{equation}
This can be proved by induction on the size of the $\mathbb{I}$ matrices $n$ (and hence the R\'{e}nyi index). The inductive hypothesis can be checked to be trivially true for $n=1$. For order $n$ the $nV_A\times nV_A$ matrix on the LHS of Eq.\ref{detstruct} can be partitioned to isolate the last $V_A\times V_A$ sub-block at the lower end. Again using the previously mentioned formula for the determinant of block matrices\cite{silvester2000determinants}, we can show that the hypothesis holds for order $n$, assuming that it is also true for order $n-1$. This completes the proof.

In our case $\hat{A}=\hat{B}=\hat{\Gamma}-2\hat{\Lambda}$. Substituting and simplifying, we get,
\begin{align}
S^{(n)}&=\frac{1}{1-n}\det\left[\left(\frac{\hat{1}+\hat{\Gamma}-2\hat{\Lambda}}{2}\right)^n+\left(\frac{\hat{1}-\hat{\Gamma}+2\hat{\Lambda}}{2}\right)^n\right]
\end{align}
which is our desired result. The analytic continuation to $n\to1$ gives us the expression for the vonNeumann entropy as quoted in Eq.\ref{SVN_final} of the main text.

\begin{acknowledgments}
 The authors are grateful
 to Sumilan Banerjee, Ahana Chakraborty and Arnab
  Sen for useful discussions and suggestions. The authors acknowledge
  the use of computational facilities at the Department of Theoretical
  Physics, Tata Institute of Fundamental Research, Mumbai for this paper.

\end{acknowledgments}

\bibliography{fermionS2}
%-----------------------------------------------------------------------------------------
\end{document}